\documentclass[aps,prd,preprint,superscriptaddress,nofootinbib]{revtex4-1}
    
    \usepackage{graphicx}  
    \usepackage{amsmath}   
    \usepackage{amssymb}   
    \usepackage{bm}        
    \usepackage{booktabs}
    \usepackage{subcaption}
    \usepackage[colorlinks=true,linkcolor=blue,urlcolor=blue,citecolor=blue]{hyperref}
    \usepackage{float}
    \usepackage{placeins}
    \usepackage{multirow}
    \usepackage{xcolor}
    \usepackage{acro}
    \usepackage{caption}
    \DeclareAcronym{pta}{
      short = PTA ,
      long = pulsar timing array ,
      short-plural-form = PTAs ,
      long-plural-form = pulsar timing arrays ,
    }

    \DeclareAcronym{sgwb}{
      short = SGWB ,
      long = stochastic gravitational-wave background ,
      short-plural-form = SGWBs ,
      long-plural-form = stochastic gravitational-wave backgrounds ,
    }

    \DeclareAcronym{orf}{
      short = ORF ,
      long = overlap reduction function ,
      short-plural-form = ORFs ,
      long-plural-form = overlap reduction functions ,
    }

    \DeclareAcronym{nanograv}{
      short = NANOGrav ,
      long = North American Nanohertz Observatory for Gravitational Waves ,
    }

    \DeclareAcronym{ska}{
      short = SKA ,
      long = Square Kilometre Array ,
    }

    \DeclareAcronym{gaianir}{
      short = Gaia-NIR ,
      long = Gaia Near Infra-Red ,
    }

    \DeclareAcronym{healpix}{
      short = HEALPix ,
      long = Hierarchical Equal Area isoLatitude Pixelation of a sphere ,
    }
    \DeclareAcronym{GR}{
      short = GR ,
      long = General Relativity ,
    }
    
\begin{document}
    		    \title{Forecasting graviton-mass constraints from the full covariance of PTA--astrometry ORF estimators}
    	\author{Jing-Hong Han}
    	\affiliation{Department of Applied Physics, College of Science, China Agricultural University, Qinghua East Road, Beijing 100083, the People's Republic of China}
    	\author{Zhi-Chao Zhao}
    	\email{zhaozc@cau.edu.cn}
    	\thanks{Corresponding author}
    	\affiliation{Department of Applied Physics, College of Science, China Agricultural University, Qinghua East Road, Beijing 100083, the People's Republic of China}
    	
    	    \begin{abstract}
       We develop a full-covariance formalism for \ac{pta}--astrometry \ac{orf} estimators and use it to forecast graviton-mass constraints from a nanohertz \ac{sgwb}. Analytic covariance expressions are derived for auto- and cross-channel \ac{orf} estimators, including signal--signal, noise--noise, and signal--noise contributions, and are validated against numerical simulations. For an observational configuration with sensitivities comparable to NANOGrav and Gaia, we obtain an expected joint 90\% upper limit of $m_g<4.41\times10^{-24}\,\mathrm{eV}/c^2$, which remains PTA-dominated and lies at the same order of magnitude as the existing NANOGrav 15-year \ac{pta}-only bound. For a future-like configuration with sensitivities comparable to the SKA and Theia/Gaia-NIR, the astrometric channels contribute significantly to the constraining power, and the joint limit improves to $m_g<0.48\times10^{-24}\,\mathrm{eV}/c^2$. These forecasts indicate that \ac{pta}--astrometry multichannel inference provides a viable avenue for improving graviton-mass constraints under next-generation observational conditions.
    	    \end{abstract}	
        
    	\maketitle
    	\acresetall
    	\section{Introduction}
    	\label{sec:intro}
    	
Whether the graviton is exactly massless is a basic question about the nature of gravity, yet the answer remains unknown~\cite{deRham2017,Will:2014kxa}. \ac{GR} predicts that gravity propagates at the speed of light and that the graviton mass is zero. This assumption underlies many calculations in gravitational physics, including gravitational-wave propagation and the evolution of large-scale structure~\cite{Will:2014kxa}. A nonzero graviton mass is not forbidden by a fundamental principle; neither gauge invariance nor Lorentz symmetry protects the graviton mass in the way they protect the photon mass~\cite{deRham2017}. The value of the graviton mass is tied to several basic properties of gravity~\cite{deRham2017,Fierz1939}. First, a nonzero graviton mass implies exponential suppression of the gravitational potential on scales set by the graviton Compton wavelength~\cite{deRham2017}, modifying gravity on cosmological scales. Second, in \ac{GR} the gravitational field is a pure spin-2 tensor field with only two transverse massless polarization degrees of freedom; once the graviton acquires a mass, longitudinal and scalar polarizations are no longer excluded by gauge symmetry, and the number of intrinsic degrees of freedom increases from two to five~\cite{deRham2017}. Massive-gravity models have also been studied in connection with the late-time accelerated expansion usually attributed to dark energy~\cite{Riess1998,Perlmutter1999,deRham2011}. Infrared corrections in quantum-gravity scenarios can likewise lead to departures from a strictly zero graviton mass~\cite{tHooft2008,deRham2017}.
Experimental bounds on the graviton mass therefore test both the range of gravity and the tensor structure of the gravitational field.
    	
Several experimental approaches constrain the graviton mass, but they differ in their model assumptions~\cite{ParticleDataGroup:2024cfk}. Yukawa-potential analyses based on planetary ephemerides ($m_g \lesssim 10^{-23}$~eV/$c^2$)~\cite{Bernus2019} and cosmological observations involving galaxy clusters and the cosmic microwave background ($m_g \lesssim 10^{-29}$~eV/$c^2$)~\cite{Desai2018,deRham2017} can yield numerically stringent limits, yet they rely on specific assumptions about the form of the gravitational-potential modification or the cosmological model, respectively. By contrast, tests based on gravitational-wave dispersion~\cite{Will1998} require comparatively few assumptions: they rely on the frequency-dependent group velocity of a massive gravitational wave, and thus depend only weakly on the detailed structure of the underlying gravity theory. The Laser Interferometer Gravitational-Wave Observatory--Virgo Collaboration has obtained $m_g \lesssim 3.09 \times 10^{-23}$~eV/$c^2$ from binary black-hole merger signals~\cite{LIGOScientific:2020tif}. Because dispersion effects are stronger at lower frequencies~\cite{Lee2010}, the nanohertz band is well suited to this test. The \ac{nanograv}, European Pulsar Timing Array/Indian Pulsar Timing Array, Parkes Pulsar Timing Array, and Chinese Pulsar Timing Array collaborations have successively reported evidence for a nanohertz \ac{sgwb}~\cite{NANOGrav2023,EPTA:2023fyk,Reardon:2023gzh,Xu:2023wog}. At these frequencies, \ac{pta} observations have improved the bound from $m_g \lesssim 10^{-22}$~eV/$c^2$ to $m_g \lesssim 10^{-24}$~eV/$c^2$ by tracking the dispersive deformation of the Hellings--Downs curve~\cite{Bernardo:2023mxc,Wu2023,Wang2024,Wu:2023rib}. Nevertheless, \acp{pta} measure only the time-delay response to the gravitational-wave signal and thus probe a single geometric response; their limited number of pulsars also restricts the achievable noise suppression~\cite{Hellings1983,Allen2023,Allen:2022dzg}. Further improvement in the nanohertz band therefore depends on complementary observables with independent geometric responses and noise sources.
    	
High-precision astrometry is one such channel~\cite{Book2011}. Unlike \acp{pta}, which measure the time delay observable $\delta z$ induced by gravitational waves, astrometric observations measure the two angular components of the apparent-position deflection, $\delta \mathbf{x}$, on the sky. The two observables are therefore different projections of the same \ac{sgwb} response~\cite{Qin2019,Book2011}. This distinction is relevant for graviton-mass studies. Subluminally propagating gravitational waves modify the \ac{pta} Hellings--Downs curve~\cite{Qin2021,Bernardo:2022rif} and the generalized \acp{orf} of astrometry~\cite{Qin2021,Mihaylov2020}; a joint analysis can use two sets of spatial correlation patterns to constrain the dispersion parameter and reduce degeneracies that are difficult to break in a single channel~\cite{Qin2021,Caliskan2024}.
    	
Astrometric searches for nanohertz gravitational waves have been studied in several recent works~\cite{Book2011,Mihaylov2018,Moore:2017ity,Klioner2018,Caliskan2024}. 
The response of stellar apparent-position deflections to an \ac{sgwb} has been derived~\cite{Book2011}, and the noise reduction obtained by cross-correlating large stellar samples, such as the $\mathcal{N}\sim 10^9$ stars observed by Gaia~\cite{Gaia2016}, has been discussed~\cite{Moore:2017ity,Klioner2018}. 
For the graviton-mass problem, analytic expressions for the generalized astrometric \acp{orf} under subluminal gravitational-wave propagation have been derived~\cite{Mihaylov2020}, joint \ac{pta}--astrometry detection in the massive-graviton framework has been explored~\cite{Qin2021}, and several studies have shown that astrometric data can improve constraints on stochastic-background parameters~\cite{Qin2019,Caliskan2024,Cruz:2024diu}. 
Here the likelihood is constructed from binned Hellings--Downs-type and \ac{orf}-type correlation-curve estimators. 
The covariance of these estimators, including both angular-bin correlations and cross-channel correlations, is therefore the central statistical object. 
Previous astrometric and joint \ac{pta}--astrometry studies have mainly formulated the relevant covariance either for signal responses and single-channel correlation analyses~\cite{Book2011,Mihaylov2018,Mihaylov2020,Caliskan2024}, or at the level of the underlying timing and astrometric observables~\cite{Cruz:2024diu} 
    	
When \ac{pta} and astrometric observations are combined in a common inference framework, their correlation-curve estimators are not statistically independent. The two classes of observations share the same realization of the \ac{sgwb}, and cross-channel estimators also share the underlying observational data. As a result, correlations arise among estimators from different observational channels. If these correlations are ignored and the single-channel likelihoods are simply multiplied together, parameter uncertainties can be underestimated and posterior constraints distorted. A covariance matrix for this problem includes the finite-source, finite-pixel realization of cosmic variance, the different noise structures of the two channels, and signal--noise coupling~\cite{Allen:2022dzg,Allen2023,Bernardo:2022xzl}. A covariance-level treatment of the joint correlation curves is therefore required for likelihood-based graviton-mass inference. With next-generation facilities such as the \ac{ska}~\cite{Janssen2015}, \ac{gaianir}~\cite{Hobbs2019}, and Theia~\cite{Theia2017}, these cross-channel correlations will need to be included in forecasts and data analyses.
    	
We develop a full-covariance framework for joint inference with \ac{pta} and astrometric correlation-curve estimators. Building on the \ac{pta} covariance treatment of Refs.~\cite{Allen:2022dzg,Allen2023}, we derive analytic covariance formulae for the joint \ac{orf} estimators, including the \ac{pta} auto-correlation, the astrometric auto-correlations, and the \ac{pta}--astrometry cross-channel estimator. These formulae include both within-channel and cross-channel covariance blocks and can be used in likelihood-based inference with parameter-dependent covariance. We validate the analytic covariance against numerical simulations and apply it to conditional Bayesian forecasts of the graviton mass in current-like and future-like observational scenarios. The resulting bounds are forecasts under the adopted assumptions rather than direct constraints from existing data.
    	
The remainder of this paper is organized as follows. Section~\ref{sec:ORFs} introduces the dispersion theory under the massive-graviton hypothesis and the generalized overlap reduction functions. Section~\ref{sec:covariance} derives the theoretical covariance formulae for the correlation-curve estimators. Section~\ref{sec:methodology} describes the all-sky simulations, the observational scenarios, and the Bayesian inference framework based on a parameterized analytic covariance. Section~\ref{sec:results} compares the analytic covariance with simulations, shows the cross-channel correlation structure, and presents the expected graviton-mass constraints under the current and future scenarios. Section~\ref{sec:conclusions} summarizes the paper and discusses limitations of the forecast.

    	\section{Gravitational Wave Observables and Overlap Reduction Functions}
    	\label{sec:ORFs}
    	
\ac{sgwb} comes from the superposition of random perturbations to the spacetime metric~\cite{Maggiore2000,Allen1996} and gives rise to two kinds of time-varying observables: shifts in pulsar signal arrival times~\cite{Detweiler1979,Sazhin1978}, described by a dimensionless time delay $\delta z(t)$, and two-dimensional angular deflections $\delta\mathbf{x}(t)$ of stellar apparent positions on the celestial sphere (measured in radians and therefore also dimensionless)~\cite{Braginsky1990,Book2011}.
The angular deflection $\delta\mathbf{x}(t)$ is resolved into two tangential components, denoted as $\delta x_{\parallel}(t)$ and $\delta x_{\perp}(t)$~\cite{Mihaylov2018}. The frequency-domain observables of a single source at frequency $f$ are collected as
\begin{equation}
\mathbf{X}(f) \equiv [\delta z(f),\; \delta x_{\parallel}(f),\; \delta x_{\perp}(f)]^T~.
\end{equation}
    	
For a Gaussian, stationary, isotropic and unpolarized \ac{sgwb}, the signal correlation of two sources located along directions $\hat{n}_A$ and $\hat{n}_B$ in the frequency domain can be expressed as~\cite{Allen1999}
    	\begin{equation}
    		\label{eq:correlation_def}
    		\langle X_A^\mu(f)\, X_B^{\nu*}(f') \rangle = \delta(f - f')\, I(f)\, \Gamma_{\mu\nu}(\gamma_{AB})~,
    	\end{equation}
where the indices $\mu,\nu \in \{z,\parallel,\perp\}$ label the three observable components, $\gamma_{AB}$ is the angular separation between sources $A$ and $B$ on the celestial sphere, and $\delta(f-f')$ reflects the statistical orthogonality of a stationary background at different frequencies. $I(f)$ is the total intensity of the \ac{sgwb}. For a power-law model with characteristic strain amplitude $A_{\rm GWB}$ and spectral index $\alpha$~\cite{Phinney2001}, we use
    	\begin{equation}
    		\label{eq:intensity_powerlaw}
    		I(f) = \frac{A_{\rm GWB}^2}{16\pi f}\left(\frac{f}{f_{\rm yr}}\right)^{2\alpha}~,
    	\end{equation}
where $f_{\rm yr} = 1\,\text{yr}^{-1}$. $\Gamma_{\mu\nu}(\gamma_{AB})$ is the generalized \ac{orf}, which depends on the observational mode and the angular separation~\cite{Christensen:1992wi,Flanagan:1993ix,Allen1999}.
    	
    	Eq.~(\ref{eq:correlation_def}) is formulated under the continuous-signal assumption. In practice, a finite observation time $T_{\rm obs}$ discretizes the continuous signal into frequency channels $f_k = k/T_{\rm obs}$ ($k = 1,2,\ldots$) with spacing $\Delta f = 1/T_{\rm obs}$. Under the independent-frequency-channel approximation, the continuous Dirac delta function transitions to a discrete Kronecker delta: $\delta(f-f') \to \delta_{kk'}/T_{\rm obs}$~\cite{Allen1999}. The discrete covariance relation for two sources in the same frequency channel $f_k$ is then~\cite{Allen1999,Allen2023}
    	    	\begin{equation}
    		\label{eq:discrete_cov}
    		\langle \mathbf{X}_A(f_k)\, \mathbf{X}_B^\dagger(f_{k'}) \rangle = \frac{1}{T_{\rm obs}}\,\delta_{kk'}\, I(f_k)\, \mathbf{C}_{AB}^{\rm loc}(f_k)~,
    	\end{equation}
where $\mathbf{C}_{AB}^{\rm loc}(f_k)$ is the $3\times 3$ correlation matrix in the local basis $\{\hat{\mathbf{e}}_z, \hat{\mathbf{e}}_{\parallel}, \hat{\mathbf{e}}_{\perp}\}$, with $\hat{\mathbf{e}}_z$ along the line of sight.
    	
A nonzero rest mass $m_g$ leaves the three-component observational structure $(z,\parallel,\perp)$ unchanged while modifying each \ac{orf} through the deviation of the gravitational-wave group velocity from the speed of light. The \ac{orf} components depend on both the angular separation and the dimensionless dispersion parameter $\epsilon(f_k)$, defined as~\cite{Mihaylov2020}
    	\begin{equation}
    		\label{eq:epsilon_def}
    		\epsilon(f_k) \equiv 1 - \frac{v_g(f_k)}{c} = 1 - \sqrt{1 - \left(\frac{m_g c^2}{h f_k}\right)^2}~,
    	\end{equation}
where $v_g(f_k)$ is the group velocity of the gravitational wave at frequency $f_k$, $h$ is the Planck constant, and $c$ is the speed of light.
    	
The local correlation matrix $\mathbf{C}_{AB}^{\rm loc}$ is~\cite{Mihaylov2018,Mihaylov2020,Qin2021}
    	\begin{equation}
    		\label{eq:C_loc}
    		\mathbf{C}_{AB}^{\rm loc}(f_k) = 
    		\begin{pmatrix}
    			\Gamma_{zz}(\gamma, \epsilon) & \Gamma_{z\parallel}(\gamma, \epsilon) & 0 \\
    			-\Gamma_{z\parallel}(\gamma, \epsilon) & \Gamma_{\parallel\parallel}(\gamma, \epsilon) & 0 \\
    			0 & 0 & \Gamma_{\perp\perp}(\gamma, \epsilon)
    		\end{pmatrix}~,
    	\end{equation}
where $\gamma \equiv \gamma_{AB}$ for brevity. The matrix elements $\Gamma_{zz}$, $\Gamma_{\parallel\parallel}$, $\Gamma_{\perp\perp}$, and $\Gamma_{z\parallel}$ denote the \ac{pta} auto-correlation, the two astrometric auto-correlations, and the \ac{pta}--astrometry cross-correlation, respectively. The cross-terms involving the perpendicular component vanish ($\Gamma_{z\perp} = \Gamma_{\parallel\perp} = 0$) as a direct consequence of parity conservation of the \ac{sgwb}~\cite{Mihaylov2018}. The minus sign in the $(2,1)$ element follows from the geometric antisymmetry of the \ac{pta}--astrometry cross-correlation: exchanging sources $A$ and $B$ reverses the projection direction of the local tangent basis $\hat{\mathbf{e}}_{\parallel}$ relative to the line joining the two points~\cite{Mihaylov2018}. In \ac{GR}, $\Gamma_{\parallel\parallel} = \Gamma_{\perp\perp}$~\cite{Qin2019}, whereas the dispersive effect of a nonzero graviton mass breaks this equality.
    	
The analytic expressions for the generalized \ac{orf} components containing the dispersion parameter $\epsilon$ are lengthy, and we refer the reader to Ref.~\cite{Mihaylov2020} for their full forms. We fixed a typo in Eq.~(16b) of the reference and checked that the fixed form would reduce to the classical forms that depend only on angular separation in the massless limit~\cite{Hellings1983,Book2011,Mihaylov2018}.

    	\section{Generalized Covariance for Cross-Channel ORF Estimators}
    	\label{sec:covariance}
    	
		When extracting the spatial correlation pattern of the \ac{sgwb} from finite, noisy data, statistical fluctuations are unavoidable~\cite{Allen2023}. In particular, once the astrometric channel and the \ac{pta}--astrometry cross-channel are incorporated into a joint analysis framework, the two observations share the same realization of the random gravitational-wave sky, and cross-channel estimators also share the underlying observational data. This gives rise to nontrivial statistical correlations among estimators from different channels, which must be described by a covariance that encompasses all channel combinations.
    	
	We label the two estimator channels by $(a, b)$ and $(c, d)$, with $a,b,c,d \in \{z, \parallel, \perp\}$. To compute their covariance, we discretize the celestial sphere into $\mathcal{N}_{\rm pix}$ pixels and bin pixel pairs by angular separation $\gamma$. Let $M_\gamma$ denote the number of pixel pairs in the $\gamma$ bin, with the $k$-th pair labeled $(i_k, j_k)$; similarly, $M_{\gamma'}$ pairs fall in the $\gamma'$ bin, and the $q$-th pair is labeled $(i_q, j_q)$, where $i$ and $j$ are pixel indices. 
The allowed pixel-pair set depends on the estimator channel. 
For the $zz$ estimator, both endpoints are restricted to pixels containing pulsars. 
For the $\parallel\parallel$ and $\perp\perp$ estimators, both endpoints are drawn from the astrometric sky pixels. 
For the cross-channel estimator $z\parallel$, the \ac{pta} endpoint is restricted to pulsar pixels, whereas the astrometric endpoint is drawn from the astrometric pixel set.
    	
		The data associated with pixel $i$ at frequency $f$, $\mathbf{D}_i(f)$, contain signal and noise,
    		$\mathbf{D}_i(f) = \mathbf{s}_i(f) + \mathbf{n}_i(f)$.
    		To simplify the subsequent derivation, we normalize the data amplitudes according to
    	\begin{equation}
    		\label{eq:normalization}
    		\mathbf{d}_i(f) \equiv \sqrt{\frac{T_{\rm obs}}{I(f)}} \, \mathbf{D}_i(f)~,
    	\end{equation}
where $T_{\rm obs}$ is the observation time span.
    	
    From Eqs.~(\ref{eq:discrete_cov}) and~(\ref{eq:normalization}), the expectation value of the signal-only normalized data product for a pixel pair $(i, j)$ with angular separation $\gamma$ is the \ac{orf}:
    $\langle d_i^a d_j^{b*} \rangle_{\rm signal} = \Gamma_{ab}(\gamma)$.
    Taking the arithmetic mean over all $M_\gamma$ pairs in the $\gamma$ bin, the \ac{orf} estimator for channel $(a, b)$ in the $\gamma$ bin is
    	\begin{equation}
    		\label{eq:estimator}
    		\hat{\Gamma}_{ab}(\gamma) = \frac{1}{M_\gamma} \sum_{k=1}^{M_\gamma} \mathrm{Re}\!\left( d_{i_k}^a \, d_{j_k}^{b*} \right) = \frac{1}{2M_\gamma} \sum_{k=1}^{M_\gamma} \left( d_{i_k}^a d_{j_k}^{b*} + d_{i_k}^{a*} d_{j_k}^b \right)~.
    	\end{equation} 
    	The covariance of $\hat{\Gamma}$ is then
    	\begin{equation*}
    	\Sigma^\Gamma_{ab,cd}(\gamma,\gamma';f)
    \equiv	\mathrm{Cov}[\hat{\Gamma}_{ab}(\gamma),\, \hat{\Gamma}_{cd}(\gamma')] = \langle \hat{\Gamma}_{ab}(\gamma)\, \hat{\Gamma}_{cd}(\gamma') \rangle - \langle \hat{\Gamma}_{ab}(\gamma) \rangle \langle \hat{\Gamma}_{cd}(\gamma') \rangle ~.
    	\end{equation*}
    	Defining the frequency-domain two-point correlation function
    	\begin{equation}
    		\label{eq:K_def}
    		K_{x,y}^{\mu\nu} \equiv \langle d_x^\mu \, d_y^{\nu*} \rangle~,
    	\end{equation}
    	one obtains, after simplification,
    	\begin{equation}
    		\label{eq:wick_result}
    		\Sigma^\Gamma_{ab,cd}(\gamma,\gamma';f)
     = \frac{1}{2M_\gamma M_{\gamma'}} \sum_{k=1}^{M_\gamma} \sum_{q=1}^{M_{\gamma'}} \mathrm{Re}\!\left[ K_{i_k,i_q}^{ac}\, K_{j_q,j_k}^{db} + K_{i_k,j_q}^{ad}\, K_{i_q,j_k}^{cb} \right]~.
    	\end{equation}
Eq.~(\ref{eq:wick_result}) applies to any pair of estimator channels through the component labels.
    	
    		The two-point correlation function $K_{x,y}^{\mu\nu}$ can be decomposed into the signal-only pair covariance $\mathcal{K}_{x,y}^{\mu\nu}$ and a noise contribution,
    	\begin{equation}
    		\label{eq:K_decomposition}
    		K_{x,y}^{\mu\nu}(f) = \mathcal{K}_{x,y}^{\mu\nu} + \frac{N_\mu(f)}{I(f)} \, \delta_{\mu\nu} \, \delta_{xy}~,
    	\end{equation}
    	where $N_\mu(f)$ is the noise power spectral density of component $\mu$, $\delta_{\mu,\nu}$ reflects the statistical independence of instrumental noise between different components, and $\delta_{xy}$ reflects the spatial independence of the noise.
    
For the \ac{pta} observable, the two-point covariance is given directly by the two sky directions. For the astrometric deflection components, however, the local $(\parallel,\perp)$ decomposition depends on the ordered pixel pair that defines the local basis. Thus the same sky pixel can acquire different local $(\parallel,\perp)$ bases in different estimator pairs. Consequently, a cross-pair covariance such as $\mathcal{K}_{i_k,j_q}$ depends not only on the sky positions of $i_k$ and $j_q$, but also on the original estimator pairs from which their local bases are inherited. Similar pair-dependent tangent-basis effects also arise in relative-astrometry \acp{orf}~\cite{Vaglio:2025tex}. To avoid notational ambiguity, we attach each local basis to its associated ordered pixel pair.

Following the geometric construction used in astrometric gravitational-wave responses~\cite{Mihaylov2018,Qin2021,Caliskan2024}, we introduce a global spherical-coordinate basis as an intermediate reference frame. For any ordered pixel pair $(x,y)$, the local basis at pixel $x$ defined by that pairing is denoted $\{\hat{\mathbf{e}}_z,\hat{\mathbf{e}}_{\parallel}^{(xy)}(x),\hat{\mathbf{e}}_{\perp}^{(xy)}(x)\}$, with $\hat{\mathbf{e}}_z$ along the line of sight, and the rotation from this local basis to the global spherical-coordinate basis $\{\hat{\mathbf{e}}_z,\hat{\mathbf{e}}_{\theta,x},\hat{\mathbf{e}}_{\phi,x}\}$ is
    \begin{equation}
    	\label{eq:M_matrix}
    	\mathbf{M}_{x}^{(xy)} =
    	\begin{pmatrix}
    		1 & 0 & 0 \\
    		0 & \hat{\mathbf{e}}_{\theta,x}\!\cdot\!\hat{\mathbf{e}}_{\parallel}^{(xy)}(x) &
    		    \hat{\mathbf{e}}_{\theta,x}\!\cdot\!\hat{\mathbf{e}}_{\perp}^{(xy)}(x) \\
    		0 & \hat{\mathbf{e}}_{\phi,x}\!\cdot\!\hat{\mathbf{e}}_{\parallel}^{(xy)}(x) &
    		    \hat{\mathbf{e}}_{\phi,x}\!\cdot\!\hat{\mathbf{e}}_{\perp}^{(xy)}(x)
    	\end{pmatrix}~.
    \end{equation}
    The superscript $(xy)$ emphasizes that even for the same pixel $x$, the local basis and the corresponding rotation matrix will in general differ whenever the original pairing changes; the \ac{pta} observable is unaffected by basis rotations.
    
    For the same ordered pixel pair $(x,y)$, the local \ac{orf} subblock $\mathbf{C}_{xy}^{\rm loc,(xy)}$ is expressed in the local basis defined by the pair $(x,y)$. The corresponding global covariance subblock is~\cite{Cruz:2024diu}
    \begin{equation}
    	\label{eq:C_glob}
    	\mathbf{C}_{xy}^{\rm glob}
    	=
    	\mathbf{M}_{x}^{(xy)}\,
    	\mathbf{C}_{xy}^{\rm loc,(xy)}\,
    	\bigl(\mathbf{M}_{y}^{(xy)}\bigr)^T~.
    \end{equation}
    This subblock depends only on the sky positions $x$ and $y$; once expressed in the global basis, it is independent of the local bases used in later projections.
    
    In the covariance summation, cross-pair subblocks are projected back from the global basis to the local bases corresponding to each estimator. For example, for
    $\boldsymbol{\mathcal{K}}_{i_k,j_q}$,
    the left endpoint $i_k$ inherits its local basis from the original pair $(i_k,j_k)$, while the right endpoint $j_q$ inherits its local basis from the original pair $(i_q,j_q)$. Therefore,
    \begin{equation}
    	\label{eq:K_projection}
    	\boldsymbol{\mathcal{K}}_{i_k,j_q}
    	=
    	\bigl(\mathbf{M}_{i_k}^{(i_k j_k)}\bigr)^T\,
    	\mathbf{C}_{i_k j_q}^{\rm glob}\,
    	\mathbf{M}_{j_q}^{(i_q j_q)}~,
    \end{equation}
    whose $(\mu,\nu)$ component is
    $\mathcal{K}_{i_k,j_q}^{\mu\nu}$.
    Similarly, $\mathcal{K}_{i_k,i_q}$, $\mathcal{K}_{j_q,j_k}$, and $\mathcal{K}_{i_q,j_k}$ are evaluated with the projection matrices corresponding to the original pairing of their respective left and right endpoints.
    	
    	Substituting Eq.~(\ref{eq:K_decomposition}) into Eq.~(\ref{eq:wick_result}), the total covariance decomposes into three terms~\cite{Allen2023}:
    	\begin{equation}
    		\label{eq:three_term}
    		\Sigma^\Gamma_{ab,cd}(\gamma,\gamma';f)
    = \mathcal{W}_{SS}^{(ab,cd)}(f) + \mathcal{W}_{NN}^{(ab,cd)}(f) + \mathcal{W}_{SN}^{(ab,cd)}(f)~.
    	\end{equation}
    	
The pure-signal term can be written as:
    	\begin{equation}
    		\label{eq:W_SS}
    		\mathcal{W}_{SS}^{(ab,cd)}(f) = \frac{1}{2M_\gamma M_{\gamma'}} \sum_{k=1}^{M_\gamma} \sum_{q=1}^{M_{\gamma'}} \mathrm{Re}\!\left[ \mathcal{K}_{i_k,i_q}^{ac}\, \mathcal{K}_{j_q,j_k}^{db} + \mathcal{K}_{i_k,j_q}^{ad}\, \mathcal{K}_{i_q,j_k}^{cb} \right]~.
    	\end{equation}
This term depends only on the celestial geometry and the \acp{orf} and contains no noise. When $a = b = c = d = z$, it reduces to the \ac{pta} Hellings--Downs form of Ref.~\cite{Allen2023}.
    	
    	The pure-noise term is nonzero only when the two pixel pairs coincide with the same or reversed ordering:
    	\begin{equation}
    		\label{eq:W_NN}
    		\mathcal{W}_{NN}^{(ab,cd)}(f) = \frac{1}{2M_\gamma M_{\gamma'}} \sum_{k=1}^{M_\gamma} \sum_{q=1}^{M_{\gamma'}} \left[ \frac{N_a N_d}{I^2} \delta_{ac}\delta_{bd}\delta_{i_k i_q}\delta_{j_k j_q} + \frac{N_a N_c}{I^2} \delta_{ad}\delta_{bc}\delta_{i_k j_q}\delta_{i_q j_k} \right]~,
    	\end{equation}
    	where the Kronecker deltas require the two channels, and the two pixel pairs, to match either in the same or in the reversed ordering.
    	
    	The signal--noise coupling term is
    	\begin{equation}
    		\label{eq:W_SN}
    		\begin{aligned}
    			\mathcal{W}_{SN}^{(ab,cd)}(f) = \frac{1}{2M_\gamma M_{\gamma'}}\sum_{k=1}^{M_\gamma} \sum_{q=1}^{M_{\gamma'}} \mathrm{Re}\bigg[
    			& \frac{N_a}{I} \delta_{ac}\delta_{i_k i_q}\, \mathcal{K}_{j_q,j_k}^{db} + \frac{N_d}{I} \delta_{bd}\delta_{j_k j_q}\, \mathcal{K}_{i_k,i_q}^{ac} \\
    			+ & \frac{N_a}{I} \delta_{ad}\delta_{i_k j_q}\, \mathcal{K}_{i_q,j_k}^{cb} + \frac{N_c}{I} \delta_{bc}\delta_{i_q j_k}\, \mathcal{K}_{i_k,j_q}^{ad} \bigg]~.
    		\end{aligned}
    	\end{equation}
    	This term depends linearly on the product of the noise power spectrum and the geometric signal projection, and is nonzero only at shared pixels.

\begin{figure}[!h]
        \centering
        \includegraphics[width=0.95\textwidth]{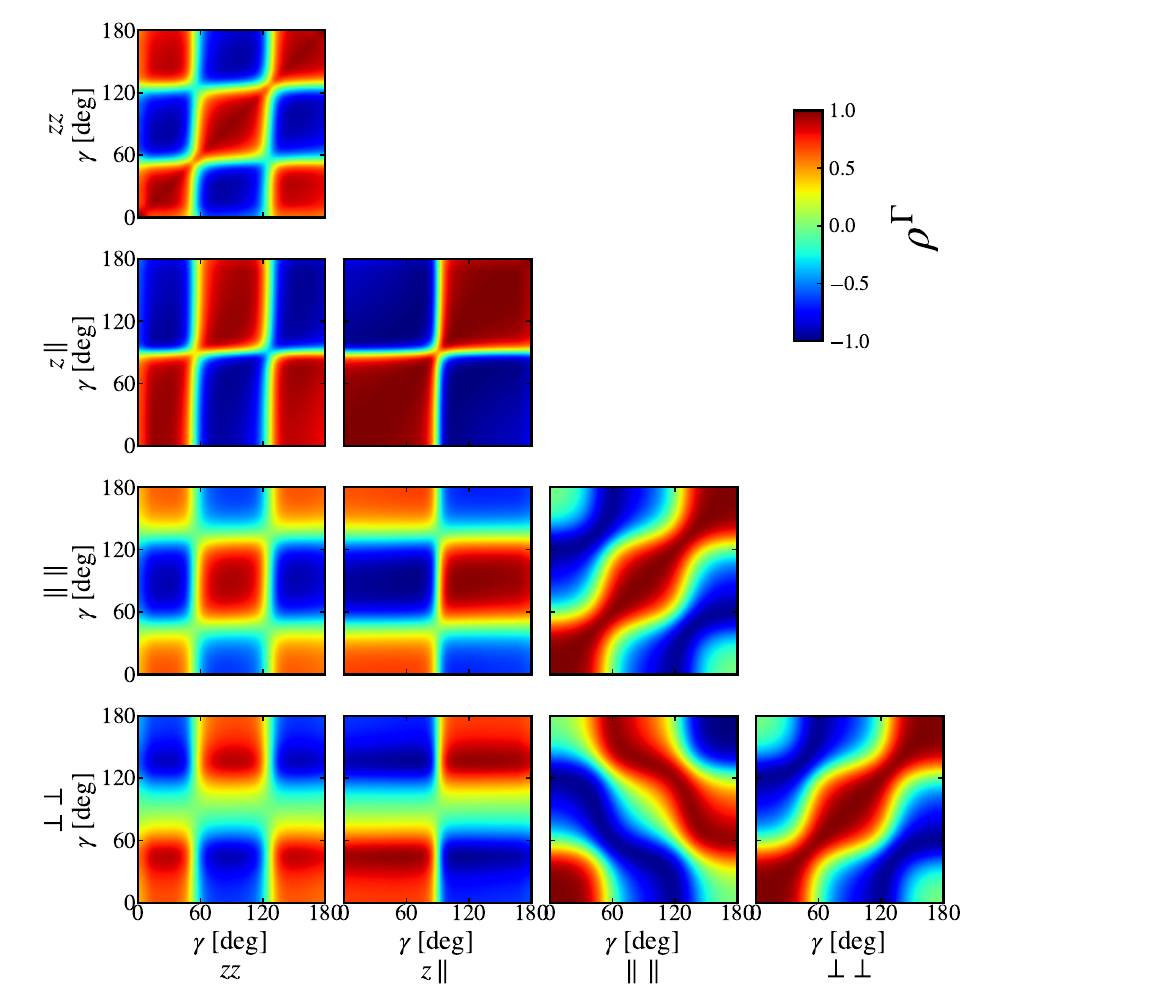}
\caption{Full correlation-coefficient matrix of the joint correlation-curve estimators in the massless limit ($m_g = 0$). The color scale corresponds to the Pearson correlation coefficient $\rho^\Gamma$ defined in Eq.~\eqref{eq:pearson_corr}.}
        \label{fig:correlation_matrix}
        \end{figure}

Fig.~\ref{fig:correlation_matrix} shows the full correlation-coefficient matrix of the joint correlation-curve estimators in the fiducial massless limit ($m_g=0$). The matrix is decomposed into $10$ lower-triangular subblocks, covering all covariance combinations among the four estimator families: the \ac{pta} channel $zz$, the \ac{pta}--astrometry cross-channel $z\parallel$, the astrometric parallel-deflection component $\parallel\parallel$, and the perpendicular-deflection component $\perp\perp$. The color shows the Pearson correlation coefficients,
\begin{equation}
\label{eq:pearson_corr}
\rho^\Gamma_{ab,cd}(\gamma_m,\gamma_n;f)
=
\frac{
\Sigma^\Gamma_{ab,cd}(\gamma_m,\gamma_n;f)
}{
\sqrt{
\Sigma^\Gamma_{ab,ab}(\gamma_m,\gamma_m;f)\,
\Sigma^\Gamma_{cd,cd}(\gamma_n,\gamma_n;f)
}
}~,
\end{equation}
where $\rho^\Gamma_{ab,cd}(\gamma_m,\gamma_n;f)$ measures the normalized correlation between $\hat{\Gamma}_{ab}(\gamma_m)$ and $\hat{\Gamma}_{cd}(\gamma_n)$. Although Fig.~\ref{fig:correlation_matrix} displays the analytic correlation matrix, the underlying covariance blocks, including those involving $z\parallel$, have also been checked against the simulation ensemble, as described in the next section.
    	
The diagonal blocks in Fig.~\ref{fig:correlation_matrix} all exhibit nonzero off-diagonal structure, indicating that estimators at different angular-separation bins are not statistically independent even within a single channel. This includes the $z\parallel$ diagonal block, where the \ac{pta}--astrometry cross-channel estimator has covariance across angular bins. The off-diagonal blocks coupling $z\parallel$ to $zz$, $\parallel\parallel$, and $\perp\perp$, as well as the blocks between the other channel pairs, are also generically nonzero. For physically distinct observables, these correlations arise primarily from the shared realization of the same \ac{sgwb}; combinations that reuse underlying observational data also receive contributions from noise-contraction terms. In the likelihood below, we retain these cross-channel covariance blocks.
    	\section{Statistical Inference Methodology}
    	\label{sec:methodology}
    	
    	\subsection{Joint Likelihood Function}
    	\label{sec:likelihood}
    		
Constraining the graviton mass requires extracting the dispersion-induced deformation of the \acp{orf} from finite, noisy data with limited sky sampling. We use Eqs.~(\ref{eq:three_term})--(\ref{eq:W_SN}) to compute the cross-channel covariance matrix entering the Bayesian likelihood.
    The statistical fluctuations of the \ac{orf} estimators depend on the parameters being constrained. We therefore fit the stochastic-background spectral parameters and the graviton mass simultaneously.
    
    In our Bayesian estimation, the free parameter set is taken to be
    	\begin{equation}
    		\label{eq:params}
    		\boldsymbol{\Theta} \equiv \{A_{\rm GWB},\; \alpha,\; m_g\}~.
    	\end{equation}
To avoid building the parameter-dependent data vector, we adopt the following data estimator for the correlation function:
\begin{equation}
	\hat{\mathcal{C}}_{ab}(\gamma, f_k)
	\equiv
	\frac{1}{M_\gamma}
	\sum_{q=1}^{M_\gamma}
	\mathrm{Re}\!\left[
		D_{i_q}^{\,a}(f_k)\,
		D_{j_q}^{\,b*}(f_k)
	\right] ~,
\end{equation}
where $(i_q,j_q)$ denotes the $q$-th pixel pair in the angular-separation bin $\gamma$. The corresponding theoretical prediction is
\begin{equation}
	\mathcal{C}^{\rm th}_{ab}(\gamma, f_k; \boldsymbol{\Theta})
	=
	\frac{I(f_k;\boldsymbol{\Theta})}{T_{\rm obs}}\,
	\Gamma_{ab}(\gamma, f_k; \boldsymbol{\Theta}) ~.
\end{equation}
To ensure consistency with the data estimator, the theoretical template is constructed using the same angular-separation binning scheme, by averaging the theoretical correlation function over all pixel pairs falling within each bin.

For a given frequency bin $f_k$ and a chosen set of channels $\mathcal{I}$, we stack the estimators across all angular-separation bins into a column data vector,
\begin{equation}
	\hat{\boldsymbol{\mathcal{C}}}_{k,\mathcal{I}}
	\equiv
	\left[
		\hat{\mathcal{C}}_{k,\mathcal{I}}(\gamma_1),\;
		\hat{\mathcal{C}}_{k,\mathcal{I}}(\gamma_2),\;
		\dots,\;
		\hat{\mathcal{C}}_{k,\mathcal{I}}(\gamma_{\mathcal{N}_{\rm bin}})
	\right]^T ~,
\end{equation}
    where $\mathcal{N}_{\rm bin}$ is the total number of angular-separation bins. The corresponding theoretical template vector is denoted
    \begin{equation}
    	\boldsymbol{\mathcal{C}}^{\rm th}_{k,\mathcal{I}}(\boldsymbol{\Theta})
    	\equiv
    	\left[
    		\mathcal{C}^{\rm th}_{k,\mathcal{I}}(\gamma_1;\boldsymbol{\Theta}),\;
    		\mathcal{C}^{\rm th}_{k,\mathcal{I}}(\gamma_2;\boldsymbol{\Theta}),\;
    		\dots,\;
    		\mathcal{C}^{\rm th}_{k,\mathcal{I}}(\gamma_{\mathcal{N}_{\rm bin}};\boldsymbol{\Theta})
    	\right]^T ~.
    \end{equation}
    
    Eqs.~(\ref{eq:three_term})--(\ref{eq:W_SN}) give the covariance of $\hat{\Gamma}$. Since
    \begin{equation}
    	\hat{\Gamma}_{ab}(\gamma,f_k;\boldsymbol{\Theta})
    	=
    	\frac{T_{\rm obs}}{I(f_k;\boldsymbol{\Theta})}\,
    	\hat{\mathcal{C}}_{ab}(\gamma,f_k) ~,
    \end{equation}
    the covariance of the fixed statistic $\hat{\mathcal{C}}$ can be written as
    \begin{equation}
    	\bm{\Sigma}^{\mathcal{C}}_{k,\mathcal{I}}(\boldsymbol{\Theta})
    	=
    	\left[
    		\frac{I(f_k;\boldsymbol{\Theta})}{T_{\rm obs}}
    	\right]^2
    	\bm{\Sigma}^{\Gamma}_{k,\mathcal{I}}(\boldsymbol{\Theta}) ~,
    \end{equation}
    where $\bm{\Sigma}^{\Gamma}_{k,\mathcal{I}}(\boldsymbol{\Theta})$ is the covariance matrix constructed from Eqs.~(\ref{eq:three_term})--(\ref{eq:W_SN}).

    The total joint log-likelihood is then given by
    \begin{equation}
    	\label{eq:likelihood}
    	\ln\mathcal{L}(\boldsymbol{\Theta}\mid \hat{\boldsymbol{\mathcal{C}}}_{\mathcal{I}})
    	=
    	-\frac{1}{2}
    	\sum_{k=1}^{\mathcal{N}_f}
    	\left\{
    		\Delta\boldsymbol{\mathcal{C}}_k^T\,
    		\bigl(\bm{\Sigma}^{\mathcal{C}}_{k,\mathcal{I}}\bigr)^{-1}
    		\Delta\boldsymbol{\mathcal{C}}_k
    		+
    		\ln\det \bm{\Sigma}^{\mathcal{C}}_{k,\mathcal{I}}
    	\right\}
    	+\mathrm{const.} ~,
    \end{equation}
    where $\mathcal{N}_f$ denotes the total number of discrete frequency bins included in the likelihood and the residual vector is
    \begin{equation}
    	\Delta\boldsymbol{\mathcal{C}}_k
    	\equiv
    	\hat{\boldsymbol{\mathcal{C}}}_{k,\mathcal{I}}
    	-
    	\boldsymbol{\mathcal{C}}^{\rm th}_{k,\mathcal{I}}(\boldsymbol{\Theta}) ~.
    \end{equation}
    
The likelihood in Eq.~(\ref{eq:likelihood}) corresponds to a single stochastic realization, so the inferred upper limit on the graviton mass fluctuates from one realization to another. A direct estimate of the median expected limit would require many simulations and a statistical summary of the resulting upper-limit distribution. To avoid that computational cost, we follow the standard high-energy-physics approach and construct an idealized fluctuation-free dataset (Asimov data) equal to the theoretical expectation~\cite{Cowan:2010js}.
    For a set of true physical parameters $\boldsymbol{\Theta}_0$ (the injected model parameters), the Asimov data for this fiducial model are
    \begin{equation}
    		\hat{\boldsymbol{\mathcal{C}}}_{k,\mathcal{I}}^{\rm Asimov} \equiv \boldsymbol{\mathcal{C}}^{\rm th}_{k,\mathcal{I}}(\boldsymbol{\Theta}_0)~.
    \end{equation}
    
Because the covariance matrix $\bm{\Sigma}_{k,\mathcal{I}}(\boldsymbol{\Theta})$ depends explicitly on the parameters, simply substituting the Asimov data into Eq.~(\ref{eq:likelihood}) would miss the ensemble-averaged contribution of the quadratic fluctuation term in the Gaussian likelihood.
    Taking the ensemble average of the multivariate Gaussian log-likelihood introduces an additional term involving the trace of the product of the true covariance $\bm{\Sigma}_{k,\mathcal{I}}(\boldsymbol{\Theta}_0)$ and the parameter-dependent inverse matrix $\bm{\Sigma}_{k,\mathcal{I}}^{-1}(\boldsymbol{\Theta})$. We therefore obtain
    \begin{equation}
    \label{eq:asimov_likelihood}
    -2 \ln\mathcal{L}_A(\boldsymbol{\Theta}) = \sum_{k=1}^{\mathcal{N}_f} \left\{ (\Delta\boldsymbol{\mathcal{C}}_{k}^{\rm A})^T\, \bigl(\bm{\Sigma}^{\mathcal{C}}_{k,\mathcal{I}}\bigr)^{-1}(\boldsymbol{\Theta})\, \Delta\boldsymbol{\mathcal{C}}_{k}^{\rm A} + \mathrm{Tr}\left[ \bigl(\bm{\Sigma}^{\mathcal{C}}_{k,\mathcal{I}}\bigr)^{-1}(\boldsymbol{\Theta}) \, \bm{\Sigma}^{\mathcal{C}}_{k,\mathcal{I}}(\boldsymbol{\Theta}_0) \right] + \ln\det \bm{\Sigma}^{\mathcal{C}}_{k,\mathcal{I}}(\boldsymbol{\Theta}) \right\}~,
\end{equation}
where $\Delta\boldsymbol{\mathcal{C}}_{k}^{\rm A} \equiv \hat{\boldsymbol{\mathcal{C}}}_{k,\mathcal{I}}^{\rm Asimov} - \boldsymbol{\mathcal{C}}^{\rm th}_{k,\mathcal{I}}(\boldsymbol{\Theta})$; the trace term compensates for the expectation introduced by the parameter dependence of the covariance.
    
    We combine the expected log-likelihood $\ln\mathcal{L}_A(\boldsymbol{\Theta})$ with prior information and perform posterior parameter sampling via Markov chain Monte Carlo (MCMC) methods. The $90\%$ posterior quantile of $m_g$ is then taken as the Asimov estimate of the median expected upper limit on the graviton mass.
    
We then perform parameter inference on the set $\{A_{\rm GWB},\alpha,m_g\}$ using the multifrequency joint likelihood. We impose Gaussian priors on $A_{\rm GWB}$ and $\alpha$, centered on the injected values,
	$A_{\rm GWB}=(2.4\pm0.6)\times10^{-15}$ and $\alpha=-2/3\pm0.3$, while imposing a uniform prior on $m_g$ in the range $0 \leq m_g \leq h f_{\min}/c^2$, where the upper bound ensures that the group velocity remains real at the lowest frequency included in the analysis. These Gaussian priors represent a future situation in which other analyses have already constrained the mean values and variances of $A_{\rm GWB}$ and $\alpha$.

    	\subsection{Observational Scenarios and Numerical Simulation}
    	\label{sec:simulation}
    	
    	To validate the analytic covariance formulae, we generate all-sky simulated data sets containing the \ac{pta}, astrometric, and cross-channel observables. Using \ac{healpix}~\cite{Gorski2004}, we discretize the celestial sphere into a finite set of pixels, denoted by $\mathcal{N}_{\rm pix}$. The fiducial model is the general-relativistic limit $m_g=0$ ($\epsilon=0$), with an injected power-law \ac{sgwb} $(A_{\rm GWB},\alpha)=(2.4\times10^{-15},-2/3)$~\cite{NANOGrav2023,Phinney2001}.
    	
At each discrete frequency $f_k$, we assemble the covariance matrix of the all-sky three-component signal field from the global-basis pairwise correlation subblocks $\mathbf{C}^{\rm glob}_{xy}$ defined by Eqs.~(\ref{eq:M_matrix}) and~(\ref{eq:C_glob}). In the present implementation, this signal covariance has dimension $3\mathcal{N}_{\rm pix}\times 3\mathcal{N}_{\rm pix}$, corresponding to one \ac{pta}-like observable and two astrometric deflection components at each sky pixel. We then perform a Cholesky decomposition, $\mathbf{C}_{\rm sig}=\mathbf{L}\mathbf{L}^\dagger$, and generate the signal sample from a standard complex Gaussian random vector $\mathbf{w}\sim\mathcal{CN}(\mathbf{0},\mathbf{I})$:
        \begin{equation}
        \label{eq:signal_gen}
        \mathbf{s}(f_k)=\sqrt{\frac{I(f_k)}{T_{\rm obs}}}\,\mathbf{L}\mathbf{w},
        \end{equation}
        where $T_{\rm obs}$ is the observation time span of the corresponding channel. The \ac{pta} data vector is constructed by restricting the \ac{pta} observable to the selected pulsar pixels, whereas the astrometric data retain the two angular-deflection components on all sky pixels. Adding the channel-specific noise $\mathbf{n}(f_k)$ then yields the simulated observational data $\mathbf{D}(f_k)=\mathbf{s}(f_k)+\mathbf{n}(f_k)$.
    	
The \ac{pta} and astrometric channels use different noise models. For the \ac{pta} channel, we include white noise from timing uncertainties and red noise from the pulsar term~\cite{Siemens2013,Moore:2014eua,Allen2023}. For astrometry, we assume that effective stars are uniformly distributed across the full sky, neglect the star term suppressed by distance geometry~\cite{Book2011,Moore:2017ity}, and retain only the position-measurement white noise~\cite{Moore:2017ity,Caliskan2024}. The frequency-domain noise for pixel $i$ in a single simulation is generated component by component as~\cite{Siemens2013,Moore:2017ity,Caliskan2024}
    	\begin{equation}
    		\label{eq:noise}
    		\begin{aligned}
    			n_{z,i}(f) &= \sqrt{\frac{N_z(f)}{T_{\rm pta}}} \; u_{z,i}~, \\
    			n_{\parallel,i}(f) &= \sqrt{\frac{N_{\parallel}(f)}{T_{\rm ast}}} \; u_{\parallel,i}~, \\
    			n_{\perp,i}(f) &= \sqrt{\frac{N_{\perp}(f)}{T_{\rm ast}}} \; u_{\perp,i}~,
    		\end{aligned}
    	\end{equation}
    	with
    	\begin{equation}
    		\label{eq:noise_psd_components}
    		\begin{aligned}
    			N_z(f) &= \Delta t_{\rm pta}\, \sigma_{\rm pta}^2\, (2\pi f)^2 + \frac{4\pi}{3}\, I(f)~, \\
    			N_{\parallel}(f) &= N_{\perp}(f) = \frac{\Delta t_{\rm ast}\, \sigma_{\rm ast}^2}{\mathcal{N}_{\rm stars}/\mathcal{N}_{\rm pix}}~.
    		\end{aligned}
    	\end{equation}
where $u_{z,i}$, $u_{\parallel,i}$, and $u_{\perp,i}$ are mutually independent draws from $\mathcal{CN}(0,1)$. $\sigma_{\rm pta}$ and $\sigma_{\rm ast}$ represent the single-source measurement precisions of the \ac{pta} and astrometric channels, respectively, and $\Delta t_{\rm pta}$ and $\Delta t_{\rm ast}$ are the respective sampling intervals. In the white-noise astrometric model, the two angular-deflection components share the same noise power spectral density, so the astrometric noise is fully determined by the pair $(n_{\parallel,i}, n_{\perp,i})$.
    	
For each data stream $r\in\{\mathrm{pta},\mathrm{ast}\}$, we define its native frequency range by $f_{\min}^{r}=1/T_r$ and $f_{\rm Nyq}^{r}=1/(2\Delta t_r)$, with native grid $f_n^r=n/T_r$.

In the single-estimator analyses, $zz$ is evaluated on the native \ac{pta} grid, while $z\parallel$, $\parallel\parallel$, and $\perp\perp$ are evaluated on the astrometric grid. This is possible in the scenarios considered here because $T_{\rm ast}\leq T_{\rm pta}$ and $\Delta t_{\rm ast}\geq\Delta t_{\rm pta}$, so the astrometric frequency range lies within the \ac{pta} frequency range. In the multichannel joint likelihood, the shared-frequency block is also evaluated on the astrometric grid and includes $zz$, $z\parallel$, $\parallel\parallel$, and $\perp\perp$ with their cross-channel covariance. Frequencies outside the astrometric range contribute only through the $zz$ estimator on the native \ac{pta} grid.

We consider two observational scenarios---current and future---whose instrumental parameter configurations are given in Tab.~\ref{tab:obs_scenarios}.
The current scenario corresponds to \ac{nanograv} 15~yr and Gaia~\cite{NANOGrav:2023ctt,Gaia2016}, while the future scenario corresponds to the \ac{ska} and Theia/\ac{gaianir}~\cite{Janssen2015,Theia2017,Hobbs2019}. 
In the numerical implementation, the current scenario is evaluated with $\mathcal{N}_{\rm pix}=768$, while the future scenario is evaluated with $\mathcal{N}_{\rm pix}=3072$. 
We verified that these resolutions are sufficient by checking that the Fisher matrix remains stable under further increases of $\mathcal{N}_{\rm pix}$.
The future scenario uses a larger pixel number because its lower astrometric noise makes the forecast more sensitive to small-scale angular information.
The quantities $\sigma_{\rm pta}$, $\mathcal{N}_{\rm psr}$, $T_{\rm pta}$, and $\Delta t_{\rm pta}$ denote the single-pulsar timing precision, the number of pulsars, the \ac{pta} observing baseline, and the \ac{pta} sampling interval, respectively. The quantities $\sigma_{\rm ast}$, $\mathcal{N}_{\rm stars}$, $T_{\rm ast}$, and $\Delta t_{\rm ast}$ denote the single-star astrometric positioning precision, the number of effective stars, the astrometric observing baseline, and the astrometric sampling interval, respectively.

    	\begin{table}[htbp]
    \centering
    \caption{Key instrumental parameter configurations for the current and future observational scenarios.}
    \label{tab:obs_scenarios}
    \resizebox{\linewidth}{!}{
    \begin{tabular}{l|c|c|c|c|c|c|c|c}
        \toprule
        \textbf{Scenario} 
        & $\sigma_{\rm pta}\,[\mathrm{ns}]$ 
        & $\mathcal{N}_{\rm psr}$ 
        & $T_{\rm pta}\,[\mathrm{yr}]$ 
        & $\Delta t_{\rm pta}\,[\mathrm{day}]$
        & $\sigma_{\rm ast}\,[\mu\mathrm{as}]$ 
        & $\mathcal{N}_{\rm stars}$ 
        & $T_{\rm ast}\,[\mathrm{yr}]$ 
        & $\Delta t_{\rm ast}\,[\mathrm{day}]$ \\
        \hline
        \midrule
        Current 
        & $80$ 
        & $50$ 
        & $15$ 
        & $14$ 
        & $100$ 
        & $1.5 \times 10^9$ 
        & $10$ 
        & $24$ \\
        Future 
        & $30$ 
        & $200$ 
        & $20$ 
        & $14$ 
        & $1$ 
        & $1.5 \times 10^9$ 
        & $20$ 
        & $24$ \\
        \bottomrule
    \end{tabular}
    }
\end{table}
    	
\begin{figure}[htbp]		
    		\centering
    		\includegraphics[width=\linewidth]{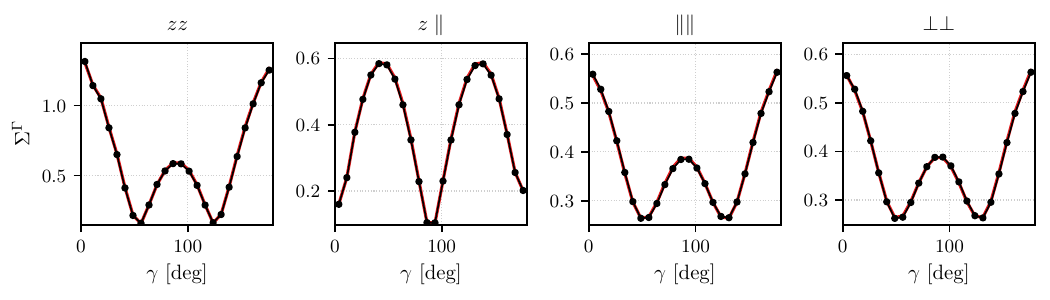}	
\caption{Comparison between empirical standard deviations and the analytic predictions from Eq.~\eqref{eq:three_term} for the fiducial injection $m_g=0$ in the future scenario. The $y$-axis represents the value of $\Sigma^\Gamma$ when $\gamma = \gamma'$. Black dots represent the empirical standard deviations obtained from $10\,000$ independent simulations, and red curves denote the theoretical standard deviations computed from the analytic covariance.}
    		\label{fig:variance_validation}
    	\end{figure}

Fig.~\ref{fig:variance_validation} shows, for the future scenario and the fiducial injection $m_g=0$ (i.e., $\epsilon=0$), the empirical standard deviations (black dots, obtained from $10\,000$ independent simulations) together with the analytic predictions (red curves) for the four estimator channels in each angular-separation bin. The empirical standard deviations are estimated directly from the simulation samples, while the theoretical curves are computed from Eqs.~(\ref{eq:three_term})--(\ref{eq:W_SN}). The current scenario was checked in the same way, but only the future case is displayed here.   	
The black dots and red curves agree across the full range of angular separations, confirming the analytic covariance against the simulation ensemble. The test covers the all-sky signal generation, noise injection, basis projection, and angular-separation binning steps. For the \ac{pta} auto-correlation channel, our results are consistent with the time-delay-correlation variance calculation of Ref.~\cite{Allen2023}. 
Fig.~\ref{fig:variance_validation} validates the diagonal elements of the covariance matrix, but we have also checked all the blocks of the covariance. The full correlation matrix shown in Fig.~\ref{fig:correlation_matrix} is also consistent with the simulation results.
    	
    	\section{Results}
    	\label{sec:results}
\begin{figure}[!htbp]
        \centering
        \includegraphics[width=0.95\textwidth]{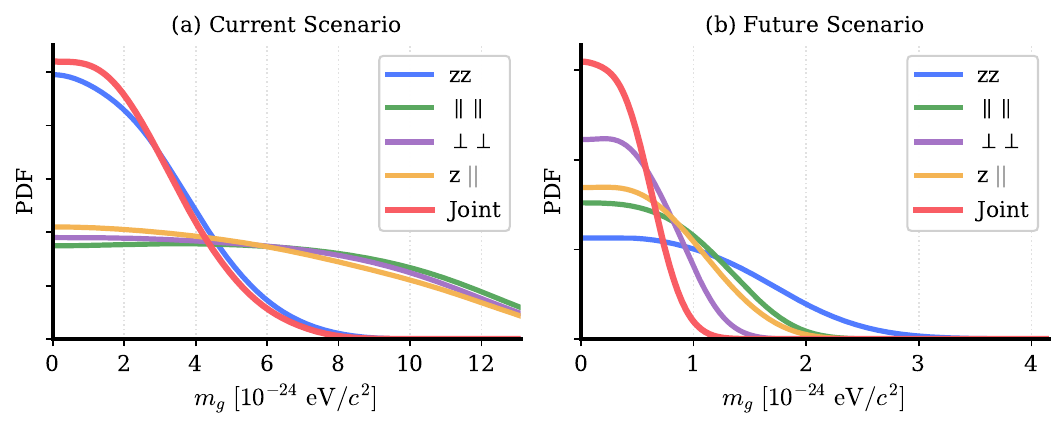}
        \caption{Marginal posterior probability density distributions of the graviton mass $m_g$. Panels (a) and (b) correspond to the current and future observational scenarios, respectively. The curves show the posterior obtained from each individual observational channel (\ac{pta} channel $zz$, cross-channel $z\parallel$, astrometric parallel-deflection component $\parallel\parallel$, and perpendicular-deflection component $\perp\perp$) and from the joint analysis.}
    \label{fig:posterior}
    \end{figure}

Fig.~\ref{fig:posterior} shows the channel-by-channel contribution to the marginalized $m_g$ posterior, and Tab.~\ref{tab:full_params_limits} reports the corresponding median expected $90\%$ upper limits on $m_g$.
    	
    	\begin{table}[htbp]
    		\centering
    		\caption{Expected $90\%$ upper limits on the graviton mass obtained from the expected log-likelihood.}
    		\label{tab:full_params_limits}
    		\renewcommand{\arraystretch}{1.15}
    		\begin{tabular}{l|c|c}
    			\toprule
    			\textbf{Channel} & \textbf{Current ($\mathrm{eV}/c^2$)} & \textbf{Future ($\mathrm{eV}/c^2$)} \\
    			\midrule
    			$zz$ & $4.68\times10^{-24}$& $1.91\times10^{-24}$\\
    			$\parallel\parallel$ & $11.1\times10^{-24}$& $1.04\times10^{-24}$\\
    			  $\perp\perp$ & $10.7\times10^{-24}$& $	0.62\times10^{-24}$\\
    			$z\parallel$ & $10.5\times10^{-24}$& $	1.08\times10^{-24}$\\
    			\textbf{Joint} & $\mathbf{4.41\times10^{-24}}$& $\mathbf{	0.48\times10^{-24}}$\\
    			\bottomrule
    		\end{tabular}
    	\end{table}
    	
In the current scenario, the constraint is dominated by the \ac{pta} channel. Fig.~\ref{fig:posterior}(a) shows that the $m_g$ posteriors from the astrometric auto-correlation and cross-channel estimators are broader and less constraining than that of the \ac{pta} channel, with $90\%$ upper limits clustering around $\sim 10^{-23}\,\mathrm{eV}/c^2$. At the current noise level, these channels contribute only marginally compared to the \ac{pta} channel. The joint analysis yields an upper limit of $4.41\times10^{-24}\,\mathrm{eV}/c^2$, remaining close to the \ac{pta}-only value of $4.68\times10^{-24}\,\mathrm{eV}/c^2$.

In the future-like scenario, the astrometric auto-correlation channels provide stronger constraints than the \ac{pta} auto-correlation channel. The $\parallel\parallel$ and $\perp\perp$ channels reach expected $90\%$ upper limits of $1.04\times10^{-24}\,\mathrm{eV}/c^2$ and $0.62\times10^{-24}\,\mathrm{eV}/c^2$, respectively, both tighter than the $zz$ result of $1.91\times10^{-24}\,\mathrm{eV}/c^2$. Among the single-channel results, the $\perp\perp$ channel gives the tightest bound. This indicates that astrometry can provide independent constraints on gravitational-wave dispersion. The $z\parallel$ cross-channel reaches $1.08\times10^{-24}\,\mathrm{eV}/c^2$, close to the $\parallel\parallel$ auto-correlation channel and tighter than the $zz$ channel, while remaining subdominant to the $\perp\perp$ channel.

The joint analysis gives a future-scenario upper limit of $0.48\times10^{-24}\,\mathrm{eV}/c^2$. The improvement relative to the single-channel bounds follows from the different geometric responses and noise properties of the \ac{pta} and astrometric observables. Note that the analysis is conditional on the Gaussian priors on $A_{\rm GWB}$ and $\alpha$. We tested the sensitivity of the $m_g$ upper limit to these priors by replacing them with uniform priors over a wide range. The resulting $90\%$ upper bound on $m_g$ changes by only about $5\%$, indicating that the graviton-mass constraint is not strongly sensitive to the assumed priors on the stochastic-background amplitude and spectral index. These forecasts indicate that \ac{pta}--astrometry multichannel inference provides a viable avenue for improving graviton-mass constraints under next-generation observational conditions.

    \section{Conclusions and Discussion}
    \label{sec:conclusions}
    
    We have presented an analytic covariance formalism for joint \ac{orf} estimators in combined pulsar-timing and astrometric analyses of \ac{sgwb}, and applied it to forecasts of graviton-mass constraints. The formalism extends the time-delay-correlation variance calculation for \acp{pta} to astrometric deflection auto-correlations and \ac{pta}--astrometry cross-channel estimators. The cross-channel blocks of the covariance matrix are generically nonzero. These covariance structures give rise to a likelihood that incorporates the full block covariance among the $zz$, $z\parallel$, $\parallel\parallel$, and $\perp\perp$ channels, accounting for correlations from the shared stochastic sky realization as well as any common underlying data.
        
	We validated the formalism against simulations in both current and future observational scenarios. The agreement between the simulated \ac{orf} estimators and the analytic prediction shows that the covariance reproduces the within-channel correlations across angular bins and the correlations between different observational channels, thereby validating both our numerical simulation scheme and the analytic formalism.
    
    In the graviton-mass forecasts, we find that the current-like scenario remains dominated by the \ac{pta} channel, and the inclusion of astrometric information yields little improvement. However, in the future-like scenario, the astrometric auto-correlation channels become individually informative, and the joint analysis gives an expected upper bound of
    $
    m_g < 4.8\times10^{-25}\,\mathrm{eV}/c^2
    $
    at $90\%$ credibility for the stated assumptions. This improvement reflects both the geometric complementarity between \ac{pta} and astrometric responses and the use of channels with distinct noise properties in the joint covariance.
    
    The parameter dependence of the covariance matrix may warrant some attention. In principle, a parameter-dependent covariance may introduce spurious information when combined with a Gaussian approximation for two-point estimators~\cite{Carron:2012pw}. In the present case, fixing the covariance to a reference model yields results close to those obtained with the parameter-dependent treatment. For the joint analysis, the resulting $m_g$ upper limits differ by only about $5\%$ in our tests.
    This issue remains for more realistic future implementations.
     
	This work shows that placing \ac{pta} and astrometric correlation estimators in a common full-covariance framework enables tests of gravitational-wave dispersion in the nanohertz band. Precision astrometry provides an independent and complementary probe alongside \ac{pta} measurements for testing whether gravity is mediated by a strictly massless field.

	The covariance structure given in this work is also relevant to other applications in which multiple observables probe the same \ac{sgwb}. The associated two-point estimators will in general be statistically correlated, and these correlations enter the inference model. Possible applications include joint studies of non-Einsteinian polarizations~\cite{Mihaylov2018}, anisotropic backgrounds~\cite{Pol2022}, or parity-violating signals~\cite{Qiao:2022mln,Gong:2021jgg,Liang:2023pbj}.

        \acknowledgments
This work is supported by the National Key Research and Development Program of China Grant No. 2021YFC2203001, the 2115 Talent Development Program of China Agricultural University, and the High-performance Computing Platform of China Agricultural University.

    	\bibliographystyle{apsrev4-1} 
    	\bibliography{refs}

\begin{thebibliography}{59}%
\makeatletter
\providecommand \@ifxundefined [1]{%
 \@ifx{#1\undefined}
}%
\providecommand \@ifnum [1]{%
 \ifnum #1\expandafter \@firstoftwo
 \else \expandafter \@secondoftwo
 \fi
}%
\providecommand \@ifx [1]{%
 \ifx #1\expandafter \@firstoftwo
 \else \expandafter \@secondoftwo
 \fi
}%
\providecommand \natexlab [1]{#1}%
\providecommand \enquote  [1]{``#1''}%
\providecommand \bibnamefont  [1]{#1}%
\providecommand \bibfnamefont [1]{#1}%
\providecommand \citenamefont [1]{#1}%
\providecommand \href@noop [0]{\@secondoftwo}%
\providecommand \href [0]{\begingroup \@sanitize@url \@href}%
\providecommand \@href[1]{\@@startlink{#1}\@@href}%
\providecommand \@@href[1]{\endgroup#1\@@endlink}%
\providecommand \@sanitize@url [0]{\catcode `\\12\catcode `\$12\catcode
  `\&12\catcode `\#12\catcode `\^12\catcode `\_12\catcode `\%12\relax}%
\providecommand \@@startlink[1]{}%
\providecommand \@@endlink[0]{}%
\providecommand \url  [0]{\begingroup\@sanitize@url \@url }%
\providecommand \@url [1]{\endgroup\@href {#1}{\urlprefix }}%
\providecommand \urlprefix  [0]{URL }%
\providecommand \Eprint [0]{\href }%
\providecommand \doibase [0]{http://dx.doi.org/}%
\providecommand \selectlanguage [0]{\@gobble}%
\providecommand \bibinfo  [0]{\@secondoftwo}%
\providecommand \bibfield  [0]{\@secondoftwo}%
\providecommand \translation [1]{[#1]}%
\providecommand \BibitemOpen [0]{}%
\providecommand \bibitemStop [0]{}%
\providecommand \bibitemNoStop [0]{.\EOS\space}%
\providecommand \EOS [0]{\spacefactor3000\relax}%
\providecommand \BibitemShut  [1]{\csname bibitem#1\endcsname}%
\let\auto@bib@innerbib\@empty
\bibitem [{\citenamefont {de~Rham}\ \emph {et~al.}(2017)\citenamefont
  {de~Rham}, \citenamefont {Deskins}, \citenamefont {Tolley},\ and\
  \citenamefont {Zhou}}]{deRham2017}%
  \BibitemOpen
  \bibfield  {author} {\bibinfo {author} {\bibfnamefont {C.}~\bibnamefont
  {de~Rham}}, \bibinfo {author} {\bibfnamefont {J.~T.}\ \bibnamefont
  {Deskins}}, \bibinfo {author} {\bibfnamefont {A.~J.}\ \bibnamefont {Tolley}},
  \ and\ \bibinfo {author} {\bibfnamefont {S.-Y.}\ \bibnamefont {Zhou}},\
  }\href {\doibase 10.1103/RevModPhys.89.025004} {\bibfield  {journal}
  {\bibinfo  {journal} {Rev. Mod. Phys.}\ }\textbf {\bibinfo {volume} {89}},\
  \bibinfo {pages} {025004} (\bibinfo {year} {2017})},\ \Eprint
  {http://arxiv.org/abs/1606.08462} {arXiv:1606.08462 [astro-ph.CO]}
  \BibitemShut {NoStop}%
\bibitem [{\citenamefont {Will}(2014)}]{Will:2014kxa}%
  \BibitemOpen
  \bibfield  {author} {\bibinfo {author} {\bibfnamefont {C.~M.}\ \bibnamefont
  {Will}},\ }\href {\doibase 10.12942/lrr-2014-4} {\bibfield  {journal}
  {\bibinfo  {journal} {Living Rev. Rel.}\ }\textbf {\bibinfo {volume} {17}},\
  \bibinfo {pages} {4} (\bibinfo {year} {2014})},\ \Eprint
  {http://arxiv.org/abs/1403.7377} {arXiv:1403.7377 [gr-qc]} \BibitemShut
  {NoStop}%
\bibitem [{\citenamefont {Fierz}\ and\ \citenamefont
  {Pauli}(1939)}]{Fierz1939}%
  \BibitemOpen
  \bibfield  {author} {\bibinfo {author} {\bibfnamefont {M.}~\bibnamefont
  {Fierz}}\ and\ \bibinfo {author} {\bibfnamefont {W.}~\bibnamefont {Pauli}},\
  }\href {\doibase 10.1098/rspa.1939.0140} {\bibfield  {journal} {\bibinfo
  {journal} {Proc. Roy. Soc. Lond. A}\ }\textbf {\bibinfo {volume} {173}},\
  \bibinfo {pages} {211} (\bibinfo {year} {1939})}\BibitemShut {NoStop}%
\bibitem [{\citenamefont {Riess}\ \emph {et~al.}(1998)\citenamefont {Riess}
  \emph {et~al.}}]{Riess1998}%
  \BibitemOpen
  \bibfield  {author} {\bibinfo {author} {\bibfnamefont {A.~G.}\ \bibnamefont
  {Riess}} \emph {et~al.} (\bibinfo {collaboration} {Supernova Search Team}),\
  }\href {\doibase 10.1086/300499} {\bibfield  {journal} {\bibinfo  {journal}
  {Astron. J.}\ }\textbf {\bibinfo {volume} {116}},\ \bibinfo {pages} {1009}
  (\bibinfo {year} {1998})},\ \Eprint {http://arxiv.org/abs/astro-ph/9805201}
  {arXiv:astro-ph/9805201} \BibitemShut {NoStop}%
\bibitem [{\citenamefont {Perlmutter}\ \emph {et~al.}(1999)\citenamefont
  {Perlmutter} \emph {et~al.}}]{Perlmutter1999}%
  \BibitemOpen
  \bibfield  {author} {\bibinfo {author} {\bibfnamefont {S.}~\bibnamefont
  {Perlmutter}} \emph {et~al.} (\bibinfo {collaboration} {Supernova Cosmology
  Project}),\ }\href {\doibase 10.1086/307221} {\bibfield  {journal} {\bibinfo
  {journal} {Astrophys. J.}\ }\textbf {\bibinfo {volume} {517}},\ \bibinfo
  {pages} {565} (\bibinfo {year} {1999})},\ \Eprint
  {http://arxiv.org/abs/astro-ph/9812133} {arXiv:astro-ph/9812133} \BibitemShut
  {NoStop}%
\bibitem [{\citenamefont {de~Rham}\ \emph {et~al.}(2011)\citenamefont
  {de~Rham}, \citenamefont {Gabadadze},\ and\ \citenamefont
  {Tolley}}]{deRham2011}%
  \BibitemOpen
  \bibfield  {author} {\bibinfo {author} {\bibfnamefont {C.}~\bibnamefont
  {de~Rham}}, \bibinfo {author} {\bibfnamefont {G.}~\bibnamefont {Gabadadze}},
  \ and\ \bibinfo {author} {\bibfnamefont {A.~J.}\ \bibnamefont {Tolley}},\
  }\href {\doibase 10.1103/PhysRevLett.106.231101} {\bibfield  {journal}
  {\bibinfo  {journal} {Phys. Rev. Lett.}\ }\textbf {\bibinfo {volume} {106}},\
  \bibinfo {pages} {231101} (\bibinfo {year} {2011})},\ \Eprint
  {http://arxiv.org/abs/1011.1232} {arXiv:1011.1232 [hep-th]} \BibitemShut
  {NoStop}%
\bibitem [{\citenamefont {'t~Hooft}(2008)}]{tHooft2008}%
  \BibitemOpen
  \bibfield  {author} {\bibinfo {author} {\bibfnamefont {G.}~\bibnamefont
  {'t~Hooft}},\ }\href {\doibase 10.1007/s10701-008-9231-3} {\bibfield
  {journal} {\bibinfo  {journal} {Found. Phys.}\ }\textbf {\bibinfo {volume}
  {38}},\ \bibinfo {pages} {733} (\bibinfo {year} {2008})},\ \Eprint
  {http://arxiv.org/abs/0804.0328} {arXiv:0804.0328 [gr-qc]} \BibitemShut
  {NoStop}%
\bibitem [{\citenamefont {Navas}\ \emph {et~al.}(2024)\citenamefont {Navas}
  \emph {et~al.}}]{ParticleDataGroup:2024cfk}%
  \BibitemOpen
  \bibfield  {author} {\bibinfo {author} {\bibfnamefont {S.}~\bibnamefont
  {Navas}} \emph {et~al.} (\bibinfo {collaboration} {Particle Data Group}),\
  }\href {\doibase 10.1103/PhysRevD.110.030001} {\bibfield  {journal} {\bibinfo
   {journal} {Phys. Rev. D}\ }\textbf {\bibinfo {volume} {110}},\ \bibinfo
  {pages} {030001} (\bibinfo {year} {2024})}\BibitemShut {NoStop}%
\bibitem [{\citenamefont {Bernus}\ \emph {et~al.}(2019)\citenamefont {Bernus},
  \citenamefont {Minazzoli}, \citenamefont {Fienga}, \citenamefont {Gastineau},
  \citenamefont {Laskar},\ and\ \citenamefont {Deram}}]{Bernus2019}%
  \BibitemOpen
  \bibfield  {author} {\bibinfo {author} {\bibfnamefont {L.}~\bibnamefont
  {Bernus}}, \bibinfo {author} {\bibfnamefont {O.}~\bibnamefont {Minazzoli}},
  \bibinfo {author} {\bibfnamefont {A.}~\bibnamefont {Fienga}}, \bibinfo
  {author} {\bibfnamefont {M.}~\bibnamefont {Gastineau}}, \bibinfo {author}
  {\bibfnamefont {J.}~\bibnamefont {Laskar}}, \ and\ \bibinfo {author}
  {\bibfnamefont {P.}~\bibnamefont {Deram}},\ }\href {\doibase
  10.1103/PhysRevLett.123.161103} {\bibfield  {journal} {\bibinfo  {journal}
  {Phys. Rev. Lett.}\ }\textbf {\bibinfo {volume} {123}},\ \bibinfo {pages}
  {161103} (\bibinfo {year} {2019})},\ \Eprint
  {http://arxiv.org/abs/1901.04307} {arXiv:1901.04307 [gr-qc]} \BibitemShut
  {NoStop}%
\bibitem [{\citenamefont {Desai}(2018)}]{Desai2018}%
  \BibitemOpen
  \bibfield  {author} {\bibinfo {author} {\bibfnamefont {S.}~\bibnamefont
  {Desai}},\ }\href {\doibase 10.1016/j.physletb.2018.01.052} {\bibfield
  {journal} {\bibinfo  {journal} {Phys. Lett. B}\ }\textbf {\bibinfo {volume}
  {778}},\ \bibinfo {pages} {325} (\bibinfo {year} {2018})},\ \Eprint
  {http://arxiv.org/abs/1708.06502} {arXiv:1708.06502 [astro-ph.CO]}
  \BibitemShut {NoStop}%
\bibitem [{\citenamefont {Will}(1998)}]{Will1998}%
  \BibitemOpen
  \bibfield  {author} {\bibinfo {author} {\bibfnamefont {C.~M.}\ \bibnamefont
  {Will}},\ }\href {\doibase 10.1103/PhysRevD.57.2061} {\bibfield  {journal}
  {\bibinfo  {journal} {Phys. Rev. D}\ }\textbf {\bibinfo {volume} {57}},\
  \bibinfo {pages} {2061} (\bibinfo {year} {1998})},\ \Eprint
  {http://arxiv.org/abs/gr-qc/9709011} {arXiv:gr-qc/9709011} \BibitemShut
  {NoStop}%
\bibitem [{\citenamefont {Abbott}\ \emph {et~al.}(2021)\citenamefont {Abbott}
  \emph {et~al.}}]{LIGOScientific:2020tif}%
  \BibitemOpen
  \bibfield  {author} {\bibinfo {author} {\bibfnamefont {R.}~\bibnamefont
  {Abbott}} \emph {et~al.} (\bibinfo {collaboration} {LIGO Scientific,
  Virgo}),\ }\href {\doibase 10.1103/PhysRevD.103.122002} {\bibfield  {journal}
  {\bibinfo  {journal} {Phys. Rev. D}\ }\textbf {\bibinfo {volume} {103}},\
  \bibinfo {pages} {122002} (\bibinfo {year} {2021})},\ \Eprint
  {http://arxiv.org/abs/2010.14529} {arXiv:2010.14529 [gr-qc]} \BibitemShut
  {NoStop}%
\bibitem [{\citenamefont {Lee}\ \emph {et~al.}(2010)\citenamefont {Lee},
  \citenamefont {Jenet}, \citenamefont {Price}, \citenamefont {Wex},\ and\
  \citenamefont {Kramer}}]{Lee2010}%
  \BibitemOpen
  \bibfield  {author} {\bibinfo {author} {\bibfnamefont {K.}~\bibnamefont
  {Lee}}, \bibinfo {author} {\bibfnamefont {F.~A.}\ \bibnamefont {Jenet}},
  \bibinfo {author} {\bibfnamefont {R.~H.}\ \bibnamefont {Price}}, \bibinfo
  {author} {\bibfnamefont {N.}~\bibnamefont {Wex}}, \ and\ \bibinfo {author}
  {\bibfnamefont {M.}~\bibnamefont {Kramer}},\ }\href {\doibase
  10.1088/0004-637X/722/2/1589} {\bibfield  {journal} {\bibinfo  {journal}
  {Astrophys. J.}\ }\textbf {\bibinfo {volume} {722}},\ \bibinfo {pages} {1589}
  (\bibinfo {year} {2010})},\ \Eprint {http://arxiv.org/abs/1008.2561}
  {arXiv:1008.2561 [astro-ph.HE]} \BibitemShut {NoStop}%
\bibitem [{\citenamefont {Agazie}\ \emph
  {et~al.}(2023{\natexlab{a}})\citenamefont {Agazie} \emph
  {et~al.}}]{NANOGrav2023}%
  \BibitemOpen
  \bibfield  {author} {\bibinfo {author} {\bibfnamefont {G.}~\bibnamefont
  {Agazie}} \emph {et~al.} (\bibinfo {collaboration} {NANOGrav}),\ }\href
  {\doibase 10.3847/2041-8213/acdac6} {\bibfield  {journal} {\bibinfo
  {journal} {Astrophys. J. Lett.}\ }\textbf {\bibinfo {volume} {951}},\
  \bibinfo {pages} {L8} (\bibinfo {year} {2023}{\natexlab{a}})},\ \Eprint
  {http://arxiv.org/abs/2306.16213} {arXiv:2306.16213 [astro-ph.HE]}
  \BibitemShut {NoStop}%
\bibitem [{\citenamefont {Antoniadis}\ \emph {et~al.}(2023)\citenamefont
  {Antoniadis} \emph {et~al.}}]{EPTA:2023fyk}%
  \BibitemOpen
  \bibfield  {author} {\bibinfo {author} {\bibfnamefont {J.}~\bibnamefont
  {Antoniadis}} \emph {et~al.} (\bibinfo {collaboration} {EPTA, InPTA}),\
  }\href {\doibase 10.1051/0004-6361/202346844} {\bibfield  {journal} {\bibinfo
   {journal} {Astron. Astrophys.}\ }\textbf {\bibinfo {volume} {678}},\
  \bibinfo {pages} {A50} (\bibinfo {year} {2023})},\ \Eprint
  {http://arxiv.org/abs/2306.16214} {arXiv:2306.16214 [astro-ph.HE]}
  \BibitemShut {NoStop}%
\bibitem [{\citenamefont {Reardon}\ \emph {et~al.}(2023)\citenamefont {Reardon}
  \emph {et~al.}}]{Reardon:2023gzh}%
  \BibitemOpen
  \bibfield  {author} {\bibinfo {author} {\bibfnamefont {D.~J.}\ \bibnamefont
  {Reardon}} \emph {et~al.},\ }\href {\doibase 10.3847/2041-8213/acdd02}
  {\bibfield  {journal} {\bibinfo  {journal} {Astrophys. J. Lett.}\ }\textbf
  {\bibinfo {volume} {951}},\ \bibinfo {pages} {L6} (\bibinfo {year} {2023})},\
  \Eprint {http://arxiv.org/abs/2306.16215} {arXiv:2306.16215 [astro-ph.HE]}
  \BibitemShut {NoStop}%
\bibitem [{\citenamefont {Xu}\ \emph {et~al.}(2023)\citenamefont {Xu} \emph
  {et~al.}}]{Xu:2023wog}%
  \BibitemOpen
  \bibfield  {author} {\bibinfo {author} {\bibfnamefont {H.}~\bibnamefont {Xu}}
  \emph {et~al.},\ }\href {\doibase 10.1088/1674-4527/acdfa5} {\bibfield
  {journal} {\bibinfo  {journal} {Res. Astron. Astrophys.}\ }\textbf {\bibinfo
  {volume} {23}},\ \bibinfo {pages} {075024} (\bibinfo {year} {2023})},\
  \Eprint {http://arxiv.org/abs/2306.16216} {arXiv:2306.16216 [astro-ph.HE]}
  \BibitemShut {NoStop}%
\bibitem [{\citenamefont {Bernardo}\ and\ \citenamefont
  {Ng}(2023{\natexlab{a}})}]{Bernardo:2023mxc}%
  \BibitemOpen
  \bibfield  {author} {\bibinfo {author} {\bibfnamefont {R.~C.}\ \bibnamefont
  {Bernardo}}\ and\ \bibinfo {author} {\bibfnamefont {K.-W.}\ \bibnamefont
  {Ng}},\ }\href {\doibase 10.1103/PhysRevD.107.L101502} {\bibfield  {journal}
  {\bibinfo  {journal} {Phys. Rev. D}\ }\textbf {\bibinfo {volume} {107}},\
  \bibinfo {pages} {L101502} (\bibinfo {year} {2023}{\natexlab{a}})},\ \Eprint
  {http://arxiv.org/abs/2302.11796} {arXiv:2302.11796 [gr-qc]} \BibitemShut
  {NoStop}%
\bibitem [{\citenamefont {Wu}\ \emph {et~al.}(2023)\citenamefont {Wu},
  \citenamefont {Chen},\ and\ \citenamefont {Huang}}]{Wu2023}%
  \BibitemOpen
  \bibfield  {author} {\bibinfo {author} {\bibfnamefont {Y.-M.}\ \bibnamefont
  {Wu}}, \bibinfo {author} {\bibfnamefont {Z.-C.}\ \bibnamefont {Chen}}, \ and\
  \bibinfo {author} {\bibfnamefont {Q.-G.}\ \bibnamefont {Huang}},\ }\href
  {\doibase 10.1103/PhysRevD.107.042003} {\bibfield  {journal} {\bibinfo
  {journal} {Phys. Rev. D}\ }\textbf {\bibinfo {volume} {107}},\ \bibinfo
  {pages} {042003} (\bibinfo {year} {2023})},\ \Eprint
  {http://arxiv.org/abs/2302.00229} {arXiv:2302.00229 [gr-qc]} \BibitemShut
  {NoStop}%
\bibitem [{\citenamefont {Wang}\ and\ \citenamefont {Zhao}(2024)}]{Wang2024}%
  \BibitemOpen
  \bibfield  {author} {\bibinfo {author} {\bibfnamefont {S.}~\bibnamefont
  {Wang}}\ and\ \bibinfo {author} {\bibfnamefont {Z.-C.}\ \bibnamefont
  {Zhao}},\ }\href {\doibase 10.1103/PhysRevD.109.L061502} {\bibfield
  {journal} {\bibinfo  {journal} {Phys. Rev. D}\ }\textbf {\bibinfo {volume}
  {109}},\ \bibinfo {pages} {L061502} (\bibinfo {year} {2024})},\ \Eprint
  {http://arxiv.org/abs/2307.04680} {arXiv:2307.04680 [astro-ph.HE]}
  \BibitemShut {NoStop}%
\bibitem [{\citenamefont {Wu}\ \emph {et~al.}(2024)\citenamefont {Wu},
  \citenamefont {Chen}, \citenamefont {Bi},\ and\ \citenamefont
  {Huang}}]{Wu:2023rib}%
  \BibitemOpen
  \bibfield  {author} {\bibinfo {author} {\bibfnamefont {Y.-M.}\ \bibnamefont
  {Wu}}, \bibinfo {author} {\bibfnamefont {Z.-C.}\ \bibnamefont {Chen}},
  \bibinfo {author} {\bibfnamefont {Y.-C.}\ \bibnamefont {Bi}}, \ and\ \bibinfo
  {author} {\bibfnamefont {Q.-G.}\ \bibnamefont {Huang}},\ }\href {\doibase
  10.1088/1361-6382/ad2a9b} {\bibfield  {journal} {\bibinfo  {journal} {Class.
  Quant. Grav.}\ }\textbf {\bibinfo {volume} {41}},\ \bibinfo {pages} {075002}
  (\bibinfo {year} {2024})},\ \Eprint {http://arxiv.org/abs/2310.07469}
  {arXiv:2310.07469 [astro-ph.CO]} \BibitemShut {NoStop}%
\bibitem [{\citenamefont {Hellings}\ and\ \citenamefont
  {Downs}(1983)}]{Hellings1983}%
  \BibitemOpen
  \bibfield  {author} {\bibinfo {author} {\bibfnamefont {R.~W.}\ \bibnamefont
  {Hellings}}\ and\ \bibinfo {author} {\bibfnamefont {G.~S.}\ \bibnamefont
  {Downs}},\ }\href {\doibase 10.1086/183954} {\bibfield  {journal} {\bibinfo
  {journal} {Astrophys. J. Lett.}\ }\textbf {\bibinfo {volume} {265}},\
  \bibinfo {pages} {L39} (\bibinfo {year} {1983})}\BibitemShut {NoStop}%
\bibitem [{\citenamefont {Allen}\ and\ \citenamefont
  {Romano}(2023)}]{Allen2023}%
  \BibitemOpen
  \bibfield  {author} {\bibinfo {author} {\bibfnamefont {B.}~\bibnamefont
  {Allen}}\ and\ \bibinfo {author} {\bibfnamefont {J.~D.}\ \bibnamefont
  {Romano}},\ }\href {\doibase 10.1103/PhysRevD.108.043026} {\bibfield
  {journal} {\bibinfo  {journal} {Phys. Rev. D}\ }\textbf {\bibinfo {volume}
  {108}},\ \bibinfo {pages} {043026} (\bibinfo {year} {2023})},\ \Eprint
  {http://arxiv.org/abs/2208.07230} {arXiv:2208.07230 [gr-qc]} \BibitemShut
  {NoStop}%
\bibitem [{\citenamefont {Allen}(2023)}]{Allen:2022dzg}%
  \BibitemOpen
  \bibfield  {author} {\bibinfo {author} {\bibfnamefont {B.}~\bibnamefont
  {Allen}},\ }\href {\doibase 10.1103/PhysRevD.107.043018} {\bibfield
  {journal} {\bibinfo  {journal} {Phys. Rev. D}\ }\textbf {\bibinfo {volume}
  {107}},\ \bibinfo {pages} {043018} (\bibinfo {year} {2023})},\ \Eprint
  {http://arxiv.org/abs/2205.05637} {arXiv:2205.05637 [gr-qc]} \BibitemShut
  {NoStop}%
\bibitem [{\citenamefont {Book}\ and\ \citenamefont
  {Flanagan}(2011)}]{Book2011}%
  \BibitemOpen
  \bibfield  {author} {\bibinfo {author} {\bibfnamefont {L.~G.}\ \bibnamefont
  {Book}}\ and\ \bibinfo {author} {\bibfnamefont {E.~E.}\ \bibnamefont
  {Flanagan}},\ }\href {\doibase 10.1103/PhysRevD.83.024024} {\bibfield
  {journal} {\bibinfo  {journal} {Phys. Rev. D}\ }\textbf {\bibinfo {volume}
  {83}},\ \bibinfo {pages} {024024} (\bibinfo {year} {2011})},\ \Eprint
  {http://arxiv.org/abs/1009.4192} {arXiv:1009.4192 [astro-ph.CO]} \BibitemShut
  {NoStop}%
\bibitem [{\citenamefont {Qin}\ \emph {et~al.}(2019)\citenamefont {Qin},
  \citenamefont {Boddy}, \citenamefont {Kamionkowski},\ and\ \citenamefont
  {Dai}}]{Qin2019}%
  \BibitemOpen
  \bibfield  {author} {\bibinfo {author} {\bibfnamefont {W.}~\bibnamefont
  {Qin}}, \bibinfo {author} {\bibfnamefont {K.~K.}\ \bibnamefont {Boddy}},
  \bibinfo {author} {\bibfnamefont {M.}~\bibnamefont {Kamionkowski}}, \ and\
  \bibinfo {author} {\bibfnamefont {L.}~\bibnamefont {Dai}},\ }\href {\doibase
  10.1103/PhysRevD.99.063002} {\bibfield  {journal} {\bibinfo  {journal} {Phys.
  Rev. D}\ }\textbf {\bibinfo {volume} {99}},\ \bibinfo {pages} {063002}
  (\bibinfo {year} {2019})},\ \Eprint {http://arxiv.org/abs/1810.02369}
  {arXiv:1810.02369 [astro-ph.CO]} \BibitemShut {NoStop}%
\bibitem [{\citenamefont {Qin}\ \emph {et~al.}(2021)\citenamefont {Qin},
  \citenamefont {Boddy},\ and\ \citenamefont {Kamionkowski}}]{Qin2021}%
  \BibitemOpen
  \bibfield  {author} {\bibinfo {author} {\bibfnamefont {W.}~\bibnamefont
  {Qin}}, \bibinfo {author} {\bibfnamefont {K.~K.}\ \bibnamefont {Boddy}}, \
  and\ \bibinfo {author} {\bibfnamefont {M.}~\bibnamefont {Kamionkowski}},\
  }\href {\doibase 10.1103/PhysRevD.103.024045} {\bibfield  {journal} {\bibinfo
   {journal} {Phys. Rev. D}\ }\textbf {\bibinfo {volume} {103}},\ \bibinfo
  {pages} {024045} (\bibinfo {year} {2021})},\ \Eprint
  {http://arxiv.org/abs/2007.11009} {arXiv:2007.11009 [gr-qc]} \BibitemShut
  {NoStop}%
\bibitem [{\citenamefont {Bernardo}\ and\ \citenamefont
  {Ng}(2023{\natexlab{b}})}]{Bernardo:2022rif}%
  \BibitemOpen
  \bibfield  {author} {\bibinfo {author} {\bibfnamefont {R.~C.}\ \bibnamefont
  {Bernardo}}\ and\ \bibinfo {author} {\bibfnamefont {K.-W.}\ \bibnamefont
  {Ng}},\ }\href {\doibase 10.1103/PhysRevD.107.044007} {\bibfield  {journal}
  {\bibinfo  {journal} {Phys. Rev. D}\ }\textbf {\bibinfo {volume} {107}},\
  \bibinfo {pages} {044007} (\bibinfo {year} {2023}{\natexlab{b}})},\ \Eprint
  {http://arxiv.org/abs/2208.12538} {arXiv:2208.12538 [gr-qc]} \BibitemShut
  {NoStop}%
\bibitem [{\citenamefont {Mihaylov}\ \emph {et~al.}(2020)\citenamefont
  {Mihaylov}, \citenamefont {Moore}, \citenamefont {Gair}, \citenamefont
  {Lasenby},\ and\ \citenamefont {Gilmore}}]{Mihaylov2020}%
  \BibitemOpen
  \bibfield  {author} {\bibinfo {author} {\bibfnamefont {D.~P.}\ \bibnamefont
  {Mihaylov}}, \bibinfo {author} {\bibfnamefont {C.~J.}\ \bibnamefont {Moore}},
  \bibinfo {author} {\bibfnamefont {J.}~\bibnamefont {Gair}}, \bibinfo {author}
  {\bibfnamefont {A.}~\bibnamefont {Lasenby}}, \ and\ \bibinfo {author}
  {\bibfnamefont {G.}~\bibnamefont {Gilmore}},\ }\href {\doibase
  10.1103/PhysRevD.101.024038} {\bibfield  {journal} {\bibinfo  {journal}
  {Phys. Rev. D}\ }\textbf {\bibinfo {volume} {101}},\ \bibinfo {pages}
  {024038} (\bibinfo {year} {2020})},\ \Eprint
  {http://arxiv.org/abs/1911.10356} {arXiv:1911.10356 [gr-qc]} \BibitemShut
  {NoStop}%
\bibitem [{\citenamefont {{\c{C}}al{\i}{\c{s}}kan}\ \emph
  {et~al.}(2024)\citenamefont {{\c{C}}al{\i}{\c{s}}kan}, \citenamefont {Chen},
  \citenamefont {Dai}, \citenamefont {Anil~Kumar}, \citenamefont {Stomberg},\
  and\ \citenamefont {Xue}}]{Caliskan2024}%
  \BibitemOpen
  \bibfield  {author} {\bibinfo {author} {\bibfnamefont {M.}~\bibnamefont
  {{\c{C}}al{\i}{\c{s}}kan}}, \bibinfo {author} {\bibfnamefont
  {Y.}~\bibnamefont {Chen}}, \bibinfo {author} {\bibfnamefont {L.}~\bibnamefont
  {Dai}}, \bibinfo {author} {\bibfnamefont {N.}~\bibnamefont {Anil~Kumar}},
  \bibinfo {author} {\bibfnamefont {I.}~\bibnamefont {Stomberg}}, \ and\
  \bibinfo {author} {\bibfnamefont {X.}~\bibnamefont {Xue}},\ }\href {\doibase
  10.1088/1475-7516/2024/05/030} {\bibfield  {journal} {\bibinfo  {journal}
  {JCAP}\ }\textbf {\bibinfo {volume} {05}},\ \bibinfo {pages} {030} (\bibinfo
  {year} {2024})},\ \Eprint {http://arxiv.org/abs/2312.03069} {arXiv:2312.03069
  [gr-qc]} \BibitemShut {NoStop}%
\bibitem [{\citenamefont {Mihaylov}\ \emph {et~al.}(2018)\citenamefont
  {Mihaylov}, \citenamefont {Moore}, \citenamefont {Gair}, \citenamefont
  {Lasenby},\ and\ \citenamefont {Gilmore}}]{Mihaylov2018}%
  \BibitemOpen
  \bibfield  {author} {\bibinfo {author} {\bibfnamefont {D.~P.}\ \bibnamefont
  {Mihaylov}}, \bibinfo {author} {\bibfnamefont {C.~J.}\ \bibnamefont {Moore}},
  \bibinfo {author} {\bibfnamefont {J.~R.}\ \bibnamefont {Gair}}, \bibinfo
  {author} {\bibfnamefont {A.}~\bibnamefont {Lasenby}}, \ and\ \bibinfo
  {author} {\bibfnamefont {G.}~\bibnamefont {Gilmore}},\ }\href {\doibase
  10.1103/PhysRevD.97.124058} {\bibfield  {journal} {\bibinfo  {journal} {Phys.
  Rev. D}\ }\textbf {\bibinfo {volume} {97}},\ \bibinfo {pages} {124058}
  (\bibinfo {year} {2018})},\ \Eprint {http://arxiv.org/abs/1804.00660}
  {arXiv:1804.00660 [gr-qc]} \BibitemShut {NoStop}%
\bibitem [{\citenamefont {Moore}\ \emph {et~al.}(2017)\citenamefont {Moore},
  \citenamefont {Mihaylov}, \citenamefont {Lasenby},\ and\ \citenamefont
  {Gilmore}}]{Moore:2017ity}%
  \BibitemOpen
  \bibfield  {author} {\bibinfo {author} {\bibfnamefont {C.~J.}\ \bibnamefont
  {Moore}}, \bibinfo {author} {\bibfnamefont {D.~P.}\ \bibnamefont {Mihaylov}},
  \bibinfo {author} {\bibfnamefont {A.}~\bibnamefont {Lasenby}}, \ and\
  \bibinfo {author} {\bibfnamefont {G.}~\bibnamefont {Gilmore}},\ }\href
  {\doibase 10.1103/PhysRevLett.119.261102} {\bibfield  {journal} {\bibinfo
  {journal} {Phys. Rev. Lett.}\ }\textbf {\bibinfo {volume} {119}},\ \bibinfo
  {pages} {261102} (\bibinfo {year} {2017})},\ \Eprint
  {http://arxiv.org/abs/1707.06239} {arXiv:1707.06239 [astro-ph.IM]}
  \BibitemShut {NoStop}%
\bibitem [{\citenamefont {Klioner}(2018)}]{Klioner2018}%
  \BibitemOpen
  \bibfield  {author} {\bibinfo {author} {\bibfnamefont {S.~A.}\ \bibnamefont
  {Klioner}},\ }\href {\doibase 10.1088/1361-6382/aa9f57} {\bibfield  {journal}
  {\bibinfo  {journal} {Class. Quant. Grav.}\ }\textbf {\bibinfo {volume}
  {35}},\ \bibinfo {pages} {045005} (\bibinfo {year} {2018})},\ \Eprint
  {http://arxiv.org/abs/1710.11474} {arXiv:1710.11474 [astro-ph.HE]}
  \BibitemShut {NoStop}%
\bibitem [{\citenamefont {Prusti}\ \emph {et~al.}(2016)\citenamefont {Prusti}
  \emph {et~al.}}]{Gaia2016}%
  \BibitemOpen
  \bibfield  {author} {\bibinfo {author} {\bibfnamefont {T.}~\bibnamefont
  {Prusti}} \emph {et~al.} (\bibinfo {collaboration} {Gaia}),\ }\href {\doibase
  10.1051/0004-6361/201629272} {\bibfield  {journal} {\bibinfo  {journal}
  {Astron. Astrophys.}\ }\textbf {\bibinfo {volume} {595}},\ \bibinfo {pages}
  {A1} (\bibinfo {year} {2016})},\ \Eprint {http://arxiv.org/abs/1609.04153}
  {arXiv:1609.04153 [astro-ph.IM]} \BibitemShut {NoStop}%
\bibitem [{\citenamefont {Cruz}\ \emph {et~al.}(2025)\citenamefont {Cruz},
  \citenamefont {Malhotra}, \citenamefont {Tasinato},\ and\ \citenamefont
  {Zavala}}]{Cruz:2024diu}%
  \BibitemOpen
  \bibfield  {author} {\bibinfo {author} {\bibfnamefont {N.~M.~J.}\
  \bibnamefont {Cruz}}, \bibinfo {author} {\bibfnamefont {A.}~\bibnamefont
  {Malhotra}}, \bibinfo {author} {\bibfnamefont {G.}~\bibnamefont {Tasinato}},
  \ and\ \bibinfo {author} {\bibfnamefont {I.}~\bibnamefont {Zavala}},\ }\href
  {\doibase 10.1103/8k1p-pzcg} {\bibfield  {journal} {\bibinfo  {journal}
  {Phys. Rev. D}\ }\textbf {\bibinfo {volume} {112}},\ \bibinfo {pages}
  {083558} (\bibinfo {year} {2025})},\ \Eprint
  {http://arxiv.org/abs/2412.14010} {arXiv:2412.14010 [astro-ph.CO]}
  \BibitemShut {NoStop}%
\bibitem [{\citenamefont {Bernardo}\ and\ \citenamefont
  {Ng}(2022)}]{Bernardo:2022xzl}%
  \BibitemOpen
  \bibfield  {author} {\bibinfo {author} {\bibfnamefont {R.~C.}\ \bibnamefont
  {Bernardo}}\ and\ \bibinfo {author} {\bibfnamefont {K.-W.}\ \bibnamefont
  {Ng}},\ }\href {\doibase 10.1088/1475-7516/2022/11/046} {\bibfield  {journal}
  {\bibinfo  {journal} {JCAP}\ }\textbf {\bibinfo {volume} {11}},\ \bibinfo
  {pages} {046} (\bibinfo {year} {2022})},\ \Eprint
  {http://arxiv.org/abs/2209.14834} {arXiv:2209.14834 [gr-qc]} \BibitemShut
  {NoStop}%
\bibitem [{\citenamefont {Janssen}\ \emph {et~al.}(2015)\citenamefont {Janssen}
  \emph {et~al.}}]{Janssen2015}%
  \BibitemOpen
  \bibfield  {author} {\bibinfo {author} {\bibfnamefont {G.}~\bibnamefont
  {Janssen}} \emph {et~al.},\ }\href {\doibase 10.22323/1.215.0037} {\bibfield
  {journal} {\bibinfo  {journal} {PoS}\ }\textbf {\bibinfo {volume}
  {AASKA14}},\ \bibinfo {pages} {037} (\bibinfo {year} {2015})},\ \Eprint
  {http://arxiv.org/abs/1501.00127} {arXiv:1501.00127 [astro-ph.IM]}
  \BibitemShut {NoStop}%
\bibitem [{\citenamefont {Hobbs}\ \emph {et~al.}(2021)\citenamefont {Hobbs}
  \emph {et~al.}}]{Hobbs2019}%
  \BibitemOpen
  \bibfield  {author} {\bibinfo {author} {\bibfnamefont {D.}~\bibnamefont
  {Hobbs}} \emph {et~al.},\ }\href {\doibase 10.1007/s10686-021-09705-z}
  {\bibfield  {journal} {\bibinfo  {journal} {Exper. Astron.}\ }\textbf
  {\bibinfo {volume} {51}},\ \bibinfo {pages} {783} (\bibinfo {year} {2021})},\
  \Eprint {http://arxiv.org/abs/1907.12535} {arXiv:1907.12535 [astro-ph.IM]}
  \BibitemShut {NoStop}%
\bibitem [{\citenamefont {Boehm}\ \emph {et~al.}(2017)\citenamefont {Boehm}
  \emph {et~al.}}]{Theia2017}%
  \BibitemOpen
  \bibfield  {author} {\bibinfo {author} {\bibfnamefont {C.}~\bibnamefont
  {Boehm}} \emph {et~al.} (\bibinfo {collaboration} {Theia}),\ }\href@noop {}
  {\  (\bibinfo {year} {2017})},\ \Eprint {http://arxiv.org/abs/1707.01348}
  {arXiv:1707.01348 [astro-ph.IM]} \BibitemShut {NoStop}%
\bibitem [{\citenamefont {Maggiore}(2000)}]{Maggiore2000}%
  \BibitemOpen
  \bibfield  {author} {\bibinfo {author} {\bibfnamefont {M.}~\bibnamefont
  {Maggiore}},\ }\href {\doibase 10.1016/S0370-1573(99)00102-7} {\bibfield
  {journal} {\bibinfo  {journal} {Phys. Rept.}\ }\textbf {\bibinfo {volume}
  {331}},\ \bibinfo {pages} {283} (\bibinfo {year} {2000})},\ \Eprint
  {http://arxiv.org/abs/gr-qc/9909001} {arXiv:gr-qc/9909001} \BibitemShut
  {NoStop}%
\bibitem [{\citenamefont {Allen}(1996)}]{Allen1996}%
  \BibitemOpen
  \bibfield  {author} {\bibinfo {author} {\bibfnamefont {B.}~\bibnamefont
  {Allen}},\ }in\ \href@noop {} {\emph {\bibinfo {booktitle} {{Les Houches
  School of Physics: Astrophysical Sources of Gravitational Radiation}}}}\
  (\bibinfo {year} {1996})\ pp.\ \bibinfo {pages} {373--417},\ \Eprint
  {http://arxiv.org/abs/gr-qc/9604033} {arXiv:gr-qc/9604033} \BibitemShut
  {NoStop}%
\bibitem [{\citenamefont {Detweiler}(1979)}]{Detweiler1979}%
  \BibitemOpen
  \bibfield  {author} {\bibinfo {author} {\bibfnamefont {S.~L.}\ \bibnamefont
  {Detweiler}},\ }\href {\doibase 10.1086/157593} {\bibfield  {journal}
  {\bibinfo  {journal} {Astrophys. J.}\ }\textbf {\bibinfo {volume} {234}},\
  \bibinfo {pages} {1100} (\bibinfo {year} {1979})}\BibitemShut {NoStop}%
\bibitem [{\citenamefont {Sazhin}(1978)}]{Sazhin1978}%
  \BibitemOpen
  \bibfield  {author} {\bibinfo {author} {\bibfnamefont {M.~V.}\ \bibnamefont
  {Sazhin}},\ }\href@noop {} {\bibfield  {journal} {\bibinfo  {journal} {Sov.
  Astron.}\ }\textbf {\bibinfo {volume} {22}},\ \bibinfo {pages} {36} (\bibinfo
  {year} {1978})}\BibitemShut {NoStop}%
\bibitem [{\citenamefont {Braginsky}\ \emph {et~al.}(1990)\citenamefont
  {Braginsky}, \citenamefont {Kardashev}, \citenamefont {Novikov},\ and\
  \citenamefont {Polnarev}}]{Braginsky1990}%
  \BibitemOpen
  \bibfield  {author} {\bibinfo {author} {\bibfnamefont {V.~B.}\ \bibnamefont
  {Braginsky}}, \bibinfo {author} {\bibfnamefont {N.~S.}\ \bibnamefont
  {Kardashev}}, \bibinfo {author} {\bibfnamefont {I.~D.}\ \bibnamefont
  {Novikov}}, \ and\ \bibinfo {author} {\bibfnamefont {A.~G.}\ \bibnamefont
  {Polnarev}},\ }\href@noop {} {\bibfield  {journal} {\bibinfo  {journal}
  {Nuovo Cim. B}\ }\textbf {\bibinfo {volume} {105}},\ \bibinfo {pages} {1141}
  (\bibinfo {year} {1990})}\BibitemShut {NoStop}%
\bibitem [{\citenamefont {Allen}\ and\ \citenamefont
  {Romano}(1999)}]{Allen1999}%
  \BibitemOpen
  \bibfield  {author} {\bibinfo {author} {\bibfnamefont {B.}~\bibnamefont
  {Allen}}\ and\ \bibinfo {author} {\bibfnamefont {J.~D.}\ \bibnamefont
  {Romano}},\ }\href {\doibase 10.1103/PhysRevD.59.102001} {\bibfield
  {journal} {\bibinfo  {journal} {Phys. Rev. D}\ }\textbf {\bibinfo {volume}
  {59}},\ \bibinfo {pages} {102001} (\bibinfo {year} {1999})},\ \Eprint
  {http://arxiv.org/abs/gr-qc/9710117} {arXiv:gr-qc/9710117} \BibitemShut
  {NoStop}%
\bibitem [{\citenamefont {Phinney}(2001)}]{Phinney2001}%
  \BibitemOpen
  \bibfield  {author} {\bibinfo {author} {\bibfnamefont {E.~S.}\ \bibnamefont
  {Phinney}},\ }\href@noop {} {\  (\bibinfo {year} {2001})},\ \Eprint
  {http://arxiv.org/abs/astro-ph/0108028} {arXiv:astro-ph/0108028} \BibitemShut
  {NoStop}%
\bibitem [{\citenamefont {Christensen}(1992)}]{Christensen:1992wi}%
  \BibitemOpen
  \bibfield  {author} {\bibinfo {author} {\bibfnamefont {N.}~\bibnamefont
  {Christensen}},\ }\href {\doibase 10.1103/PhysRevD.46.5250} {\bibfield
  {journal} {\bibinfo  {journal} {Phys. Rev. D}\ }\textbf {\bibinfo {volume}
  {46}},\ \bibinfo {pages} {5250} (\bibinfo {year} {1992})}\BibitemShut
  {NoStop}%
\bibitem [{\citenamefont {Flanagan}(1993)}]{Flanagan:1993ix}%
  \BibitemOpen
  \bibfield  {author} {\bibinfo {author} {\bibfnamefont {E.~E.}\ \bibnamefont
  {Flanagan}},\ }\href {\doibase 10.1103/PhysRevD.48.2389} {\bibfield
  {journal} {\bibinfo  {journal} {Phys. Rev. D}\ }\textbf {\bibinfo {volume}
  {48}},\ \bibinfo {pages} {2389} (\bibinfo {year} {1993})},\ \Eprint
  {http://arxiv.org/abs/astro-ph/9305029} {arXiv:astro-ph/9305029} \BibitemShut
  {NoStop}%
\bibitem [{\citenamefont {Vaglio}\ \emph {et~al.}(2025)\citenamefont {Vaglio},
  \citenamefont {Falxa}, \citenamefont {Mentasti}, \citenamefont {Renzini},
  \citenamefont {Kuntz}, \citenamefont {Barausse}, \citenamefont {Contaldi},\
  and\ \citenamefont {Sesana}}]{Vaglio:2025tex}%
  \BibitemOpen
  \bibfield  {author} {\bibinfo {author} {\bibfnamefont {M.}~\bibnamefont
  {Vaglio}}, \bibinfo {author} {\bibfnamefont {M.}~\bibnamefont {Falxa}},
  \bibinfo {author} {\bibfnamefont {G.}~\bibnamefont {Mentasti}}, \bibinfo
  {author} {\bibfnamefont {A.~I.}\ \bibnamefont {Renzini}}, \bibinfo {author}
  {\bibfnamefont {A.}~\bibnamefont {Kuntz}}, \bibinfo {author} {\bibfnamefont
  {E.}~\bibnamefont {Barausse}}, \bibinfo {author} {\bibfnamefont
  {C.}~\bibnamefont {Contaldi}}, \ and\ \bibinfo {author} {\bibfnamefont
  {A.}~\bibnamefont {Sesana}},\ }\href@noop {} {\  (\bibinfo {year} {2025})},\
  \Eprint {http://arxiv.org/abs/2507.18593} {arXiv:2507.18593 [gr-qc]}
  \BibitemShut {NoStop}%
\bibitem [{\citenamefont {Cowan}\ \emph {et~al.}(2011)\citenamefont {Cowan},
  \citenamefont {Cranmer}, \citenamefont {Gross},\ and\ \citenamefont
  {Vitells}}]{Cowan:2010js}%
  \BibitemOpen
  \bibfield  {author} {\bibinfo {author} {\bibfnamefont {G.}~\bibnamefont
  {Cowan}}, \bibinfo {author} {\bibfnamefont {K.}~\bibnamefont {Cranmer}},
  \bibinfo {author} {\bibfnamefont {E.}~\bibnamefont {Gross}}, \ and\ \bibinfo
  {author} {\bibfnamefont {O.}~\bibnamefont {Vitells}},\ }\href {\doibase
  10.1140/epjc/s10052-011-1554-0} {\bibfield  {journal} {\bibinfo  {journal}
  {Eur. Phys. J. C}\ }\textbf {\bibinfo {volume} {71}},\ \bibinfo {pages}
  {1554} (\bibinfo {year} {2011})},\ \bibinfo {note} {[Erratum: Eur.Phys.J.C
  73, 2501 (2013)]},\ \Eprint {http://arxiv.org/abs/1007.1727} {arXiv:1007.1727
  [physics.data-an]} \BibitemShut {NoStop}%
\bibitem [{\citenamefont {G{\'o}rski}\ \emph {et~al.}(2005)\citenamefont
  {G{\'o}rski}, \citenamefont {Hivon}, \citenamefont {Banday}, \citenamefont
  {Wandelt}, \citenamefont {Hansen}, \citenamefont {Reinecke},\ and\
  \citenamefont {Bartelman}}]{Gorski2004}%
  \BibitemOpen
  \bibfield  {author} {\bibinfo {author} {\bibfnamefont {K.~M.}\ \bibnamefont
  {G{\'o}rski}}, \bibinfo {author} {\bibfnamefont {E.}~\bibnamefont {Hivon}},
  \bibinfo {author} {\bibfnamefont {A.~J.}\ \bibnamefont {Banday}}, \bibinfo
  {author} {\bibfnamefont {B.~D.}\ \bibnamefont {Wandelt}}, \bibinfo {author}
  {\bibfnamefont {F.~K.}\ \bibnamefont {Hansen}}, \bibinfo {author}
  {\bibfnamefont {M.}~\bibnamefont {Reinecke}}, \ and\ \bibinfo {author}
  {\bibfnamefont {M.}~\bibnamefont {Bartelman}},\ }\href {\doibase
  10.1086/427976} {\bibfield  {journal} {\bibinfo  {journal} {Astrophys. J.}\
  }\textbf {\bibinfo {volume} {622}},\ \bibinfo {pages} {759} (\bibinfo {year}
  {2005})},\ \Eprint {http://arxiv.org/abs/astro-ph/0409513}
  {arXiv:astro-ph/0409513} \BibitemShut {NoStop}%
\bibitem [{\citenamefont {Siemens}\ \emph {et~al.}(2013)\citenamefont
  {Siemens}, \citenamefont {Ellis}, \citenamefont {Jenet},\ and\ \citenamefont
  {Romano}}]{Siemens2013}%
  \BibitemOpen
  \bibfield  {author} {\bibinfo {author} {\bibfnamefont {X.}~\bibnamefont
  {Siemens}}, \bibinfo {author} {\bibfnamefont {J.}~\bibnamefont {Ellis}},
  \bibinfo {author} {\bibfnamefont {F.}~\bibnamefont {Jenet}}, \ and\ \bibinfo
  {author} {\bibfnamefont {J.~D.}\ \bibnamefont {Romano}},\ }\href {\doibase
  10.1088/0264-9381/30/22/224015} {\bibfield  {journal} {\bibinfo  {journal}
  {Class. Quant. Grav.}\ }\textbf {\bibinfo {volume} {30}},\ \bibinfo {pages}
  {224015} (\bibinfo {year} {2013})},\ \Eprint {http://arxiv.org/abs/1305.3196}
  {arXiv:1305.3196 [astro-ph.IM]} \BibitemShut {NoStop}%
\bibitem [{\citenamefont {Moore}\ \emph {et~al.}(2015)\citenamefont {Moore},
  \citenamefont {Taylor},\ and\ \citenamefont {Gair}}]{Moore:2014eua}%
  \BibitemOpen
  \bibfield  {author} {\bibinfo {author} {\bibfnamefont {C.~J.}\ \bibnamefont
  {Moore}}, \bibinfo {author} {\bibfnamefont {S.~R.}\ \bibnamefont {Taylor}}, \
  and\ \bibinfo {author} {\bibfnamefont {J.~R.}\ \bibnamefont {Gair}},\ }\href
  {\doibase 10.1088/0264-9381/32/5/055004} {\bibfield  {journal} {\bibinfo
  {journal} {Class. Quant. Grav.}\ }\textbf {\bibinfo {volume} {32}},\ \bibinfo
  {pages} {055004} (\bibinfo {year} {2015})},\ \Eprint
  {http://arxiv.org/abs/1406.5199} {arXiv:1406.5199 [astro-ph.IM]} \BibitemShut
  {NoStop}%
\bibitem [{\citenamefont {Agazie}\ \emph
  {et~al.}(2023{\natexlab{b}})\citenamefont {Agazie} \emph
  {et~al.}}]{NANOGrav:2023ctt}%
  \BibitemOpen
  \bibfield  {author} {\bibinfo {author} {\bibfnamefont {G.}~\bibnamefont
  {Agazie}} \emph {et~al.} (\bibinfo {collaboration} {NANOGrav}),\ }\href
  {\doibase 10.3847/2041-8213/acda88} {\bibfield  {journal} {\bibinfo
  {journal} {Astrophys. J. Lett.}\ }\textbf {\bibinfo {volume} {951}},\
  \bibinfo {pages} {L10} (\bibinfo {year} {2023}{\natexlab{b}})},\ \Eprint
  {http://arxiv.org/abs/2306.16218} {arXiv:2306.16218 [astro-ph.HE]}
  \BibitemShut {NoStop}%
\bibitem [{\citenamefont {Carron}(2013)}]{Carron:2012pw}%
  \BibitemOpen
  \bibfield  {author} {\bibinfo {author} {\bibfnamefont {J.}~\bibnamefont
  {Carron}},\ }\href {\doibase 10.1051/0004-6361/201220538} {\bibfield
  {journal} {\bibinfo  {journal} {Astron. Astrophys.}\ }\textbf {\bibinfo
  {volume} {551}},\ \bibinfo {pages} {A88} (\bibinfo {year} {2013})},\ \Eprint
  {http://arxiv.org/abs/1204.4724} {arXiv:1204.4724 [astro-ph.CO]} \BibitemShut
  {NoStop}%
\bibitem [{\citenamefont {Pol}\ \emph {et~al.}(2022)\citenamefont {Pol},
  \citenamefont {Taylor},\ and\ \citenamefont {Romano}}]{Pol2022}%
  \BibitemOpen
  \bibfield  {author} {\bibinfo {author} {\bibfnamefont {N.}~\bibnamefont
  {Pol}}, \bibinfo {author} {\bibfnamefont {S.~R.}\ \bibnamefont {Taylor}}, \
  and\ \bibinfo {author} {\bibfnamefont {J.~D.}\ \bibnamefont {Romano}},\
  }\href {\doibase 10.3847/1538-4357/ac9836} {\bibfield  {journal} {\bibinfo
  {journal} {Astrophys. J.}\ }\textbf {\bibinfo {volume} {940}},\ \bibinfo
  {pages} {173} (\bibinfo {year} {2022})},\ \Eprint
  {http://arxiv.org/abs/2206.09936} {arXiv:2206.09936 [astro-ph.HE]}
  \BibitemShut {NoStop}%
\bibitem [{\citenamefont {Qiao}\ \emph {et~al.}(2023)\citenamefont {Qiao},
  \citenamefont {Li}, \citenamefont {Zhu}, \citenamefont {Ji}, \citenamefont
  {Li},\ and\ \citenamefont {Zhao}}]{Qiao:2022mln}%
  \BibitemOpen
  \bibfield  {author} {\bibinfo {author} {\bibfnamefont {J.}~\bibnamefont
  {Qiao}}, \bibinfo {author} {\bibfnamefont {Z.}~\bibnamefont {Li}}, \bibinfo
  {author} {\bibfnamefont {T.}~\bibnamefont {Zhu}}, \bibinfo {author}
  {\bibfnamefont {R.}~\bibnamefont {Ji}}, \bibinfo {author} {\bibfnamefont
  {G.}~\bibnamefont {Li}}, \ and\ \bibinfo {author} {\bibfnamefont
  {W.}~\bibnamefont {Zhao}},\ }\href {\doibase 10.3389/fspas.2022.1109086}
  {\bibfield  {journal} {\bibinfo  {journal} {Front. Astron. Space Sci.}\
  }\textbf {\bibinfo {volume} {9}},\ \bibinfo {pages} {1109086} (\bibinfo
  {year} {2023})},\ \Eprint {http://arxiv.org/abs/2211.16825} {arXiv:2211.16825
  [gr-qc]} \BibitemShut {NoStop}%
\bibitem [{\citenamefont {Gong}\ \emph {et~al.}(2022)\citenamefont {Gong},
  \citenamefont {Zhu}, \citenamefont {Niu}, \citenamefont {Wu}, \citenamefont
  {Cui}, \citenamefont {Zhang}, \citenamefont {Zhao},\ and\ \citenamefont
  {Wang}}]{Gong:2021jgg}%
  \BibitemOpen
  \bibfield  {author} {\bibinfo {author} {\bibfnamefont {C.}~\bibnamefont
  {Gong}}, \bibinfo {author} {\bibfnamefont {T.}~\bibnamefont {Zhu}}, \bibinfo
  {author} {\bibfnamefont {R.}~\bibnamefont {Niu}}, \bibinfo {author}
  {\bibfnamefont {Q.}~\bibnamefont {Wu}}, \bibinfo {author} {\bibfnamefont
  {J.-L.}\ \bibnamefont {Cui}}, \bibinfo {author} {\bibfnamefont
  {X.}~\bibnamefont {Zhang}}, \bibinfo {author} {\bibfnamefont
  {W.}~\bibnamefont {Zhao}}, \ and\ \bibinfo {author} {\bibfnamefont
  {A.}~\bibnamefont {Wang}},\ }\href {\doibase 10.1103/PhysRevD.105.044034}
  {\bibfield  {journal} {\bibinfo  {journal} {Phys. Rev. D}\ }\textbf {\bibinfo
  {volume} {105}},\ \bibinfo {pages} {044034} (\bibinfo {year} {2022})},\
  \Eprint {http://arxiv.org/abs/2112.06446} {arXiv:2112.06446 [gr-qc]}
  \BibitemShut {NoStop}%
\bibitem [{\citenamefont {Liang}\ \emph {et~al.}(2024)\citenamefont {Liang},
  \citenamefont {Lin}, \citenamefont {Trodden},\ and\ \citenamefont
  {Wong}}]{Liang:2023pbj}%
  \BibitemOpen
  \bibfield  {author} {\bibinfo {author} {\bibfnamefont {Q.}~\bibnamefont
  {Liang}}, \bibinfo {author} {\bibfnamefont {M.-X.}\ \bibnamefont {Lin}},
  \bibinfo {author} {\bibfnamefont {M.}~\bibnamefont {Trodden}}, \ and\
  \bibinfo {author} {\bibfnamefont {S.~S.~C.}\ \bibnamefont {Wong}},\ }\href
  {\doibase 10.1103/PhysRevD.109.083028} {\bibfield  {journal} {\bibinfo
  {journal} {Phys. Rev. D}\ }\textbf {\bibinfo {volume} {109}},\ \bibinfo
  {pages} {083028} (\bibinfo {year} {2024})},\ \Eprint
  {http://arxiv.org/abs/2309.16666} {arXiv:2309.16666 [astro-ph.CO]}
  \BibitemShut {NoStop}%
\end{thebibliography}%
    	
    \end{document}